\documentclass[aip, amsmath,amssymb,reprint]{revtex4-1}
\usepackage{color}
\usepackage{graphicx}
\newcommand{\rev}[1]{{\color{black}#1}}

\usepackage{accents}
\renewcommand*{\dot}[1]{%
	\accentset{\mbox{\large\bfseries .}}{#1}}

\let\eps\varepsilon
\newcommand{\R}{\mathbb R}
\newcommand{\bA}{\mathbf A}

\newcommand{\bD}{\mathbf D}

\newcommand{\bH}{\mathbf H}
\newcommand{\bI}{\mathbf I}

\newcommand{\bP}{\mathbf P}

\newcommand{\ba}{\mathbf a}

\newcommand{\bb}{\mathbf b}

\newcommand{\bn}{\mathbf n}
\newcommand{\be}{\mathbf e}

\newcommand{\bu}{\mathbf u}
\newcommand{\bv}{\mathbf v}

\newcommand{\bx}{\mathbf x}
\newcommand{\by}{\mathbf y}
\newcommand{\bz}{\mathbf z}

\newcommand{\I}{\mathcal I}

\newcommand{\T}{\mathcal T}

\newcommand{\Div}{\mathop{\rm div}}
\newcommand{\divG}{{\mathop{\,\rm div}}_{\Gamma}}
\newcommand{\gradG}{\nabla_{\Gamma}}

\renewcommand{\div}{\mathop{\rm div}}

\newcommand{\tr}{{\rm tr}}

 \newtheorem{rem}{Remark}

\begin{document}

\title{On equilibrium states of fluid membranes}


\author{Maxim A. Olshanskii}
\email[]{maolshanskiy@uh.edu}
\homepage[]{www.math.uh.edu/~molshan}
\affiliation{Department of Mathematics, University of Houston, 3551 Cullen Blvd, Houston, Texas 77204, USA}


\date{\today}

\begin{abstract}
The paper studies the equilibrium configurations of inextensible elastic membranes exhibiting lateral fluidity.
 Using a continuum description of the membrane's motions based on the surface Navier--Stokes equations with bending forces, the paper derives differential equations governing the mechanical equilibrium. The equilibrium conditions are found to be independent of lateral viscosity and relate tension, pressure, and tangential velocity of the fluid.
 These conditions suggest that \rev{either the lateral fluid motion ceases or non-decaying stationary flow of mass can only be supported by surfaces with Killing vector fields, such as axisymmetric shapes.} 
  A shape equation is derived that extends the classical Helfrich model with an area constraint to membranes of non-negligible  mass.  Furthermore, the paper suggests a simple numerical method to compute solutions of the shape equation. Numerical experiments conducted reveal a diverse family of equilibrium configurations. \rev{The stability of equilibrium states involving lateral flow of mass remains an unresolved question.}
\end{abstract}


\rightline{\footnotesize preprint accepted to PoF}
\maketitle 

\section{Introduction}
Motivated by applications in cell biology, there has been extensive research on studying equilibrium configurations of fluid membranes, their stability, and transformations\cite{deuling1976curvature,jenkins1977static,peterson1985instability,seifert1991shape,seifert1995morphology,seifert1997configurations,deserno2015fluid}. A now-classical energetic approach to describe the statics and dynamics of fluid membranes was pioneered by Canham~\cite{canham1970minimum} and Helfrich~\cite{helfrich1973elastic}. According to the Canham--Helfrich theory, an equilibrium shape of a membrane minimizes a curvature energy functional subject to possible constraints.

More complex models account for in-plane fluidity exhibited by the membranes. In a continuum-based modeling approach, the membrane is represented by a \emph{material surface} that supports a density flow and may deform driven by both elastic and hydrodynamic forces. The development and analysis {of continuum-based models and their application to} numerical simulation of fluid membrane dynamics is an area of active research~\cite{hu2007continuum,arroyo2009,rangamani2013interaction,barrett2015numerical,jankuhn2018incompressible,nitschke2019hydrodynamic,voigt2019fluid,reuther2020numerical,sahu2020arbitrary,krause2022numerical}. In Ref.\cite{torres2019modelling}, such models of elastic fluid thin sheets were given the name \emph{fluid deformable surfaces}.

A system of equations governing the motion of {fluid deformable surfaces} consists of the surface Navier-Stokes equations coupled with an elasticity model and posed on a time-dependent surface, while the surface evolution is defined by the hydrodynamic part of the solution; see Sec.~\ref{sectmodel} for further details. The governing equations represent the conservation of momentum so that a steady state of the system is a mechanical equilibrium. Existing research on {fluid deformable} surfaces mainly addresses the system evolution, {with only a few papers addressing the problem numerically \cite{rangamani2013interaction,rodrigues2015semi,torres2019modelling,reuther2020numerical,krause2022numerical}, among which Refs.\cite{reuther2020numerical,krause2022numerical} considered relaxation to equilibrium.} In the present study, we are interested in equilibrium configurations. 

Assuming a steady state solution to the surface Navier--Stokes equations with vanishing external lateral forces, we deduce three conditions for such solutions to exist: the radial motions are zero, the lateral motions correspond to a Killing field on manifold, and the third condition requires that a specific surface pressure (defined in \eqref{pressure}) is constant.
The first two conditions imply that only two scenarios of equilibrium are possible: Either the fluid motion ceases and the problem reduces to the well studied one of finding a shape of minimal curvature energy under the area constraint, or an additional geometrical constraint arises for the equilibrium shape to support a non-decaying lateral fluid flow. The constraint is satisfied by axisymmetric shapes.

To explore the second scenario, we utilize the third condition and a specific elasticity model (the simplest Helfrich model in this paper) to derive the \emph{shape equation}, which is an equation satisfied by geometrical quantities of the system in equilibrium. We introduce a numerical approach to solve the shape equation, which takes advantage of the axial symmetry of the unknown surface. We solve the shape equation numerically to obtain branches of shapes with a fixed surface area and varying interior volume for several sets of physical parameters. In particular, we find steady states of the surface fluid equations with vanishing elastic forces, which correspond to equilibrium configurations of `pure fluidic' membranes. \rev{The stability of the discovered equilibrium states remains an open question.}

The remainder of the paper is organized into four sections. Section~\ref{sectmodel} reviews the continuum model of a fluid-elastic membrane and derives the equilibrium conditions and the shape equation. Section~\ref{sec:num} introduces a numerical solver. The computed shapes are discussed in Section~\ref{sec:shapes}. Section~\ref{sec:concl} provides a few concluding remarks.

 For the model setup and analysis, this paper employs elementary tangential calculus~\cite{delfour2000tangential,jankuhn2018incompressible} in embedding $\R^3$ space, thus avoiding any calculations in local surface coordinates.

 \section{A deforming fluid--elastic membrane} \label{sectmodel}
Following the continuum-mechanical description, we represent the membrane as a smooth closed time-dependent surface $\Gamma(t) \subset \mathbb{R}^3$ with a density distribution $\rho(x,t)$~\cite{GurtinMurdoch75,MurdochCohen79}. Let $\bu$ be a smooth velocity field of the density flow on $\Gamma$, i.e., $\bu(x,t)$ is the velocity of the material point $x \in \Gamma(t)$. In general, $\bu$ is not necessarily tangential to $\Gamma$, and its normal component defines the geometric evolution of $\Gamma$.

To formulate equations governing the motion of the membrane, we need a few surface quantities and tangential differential operators. Let $\bn$ be an outward-pointing normal vector on $\Gamma$, and let $\bP = \bI - \bn\bn^T$ denote the normal projector. The surface gradient $\nabla_\Gamma p$ of a scalar function $p:\Gamma \to \mathbb{R}$ can be defined as $\nabla_\Gamma p = \bP \nabla p^e$, where $p^e$ is an arbitrary smooth extension of $p$ in a neighborhood of $\Gamma$. Also, $\nabla_\Gamma \bu = \bP(\nabla_\Gamma u_1, \nabla_\Gamma u_2, \nabla_\Gamma u_3)^T$ is a surface gradient of a vector field $\bu = (u_1, u_2, u_3)^T:\Gamma \to \mathbb{R}^3$. In other words, $\nabla_\Gamma \bu$ is a covariant gradient if $\bu$ is tangential to $\Gamma$, $\divG \bu = \mathrm{tr}(\gradG \bu)$ is the surface divergence, and $\Delta_\Gamma p = \divG \nabla_\Gamma p$ is the Laplace-Beltrami operator. For a tensor field $\bA = [\ba_1, \ba_2, \ba_3]: \Gamma \to \mathbb{R}^{3\times 3}$, the surface divergence $\divG \bA$ is defined row-wise.

We assume an inextensible viscous membrane and use the Boussinesq–Scriven constitutive relation for the surface stress tensor. Conservation of mass and linear momentum for an arbitrary material area $\gamma(t)\subset\Gamma(t)$ leads to the evolving surface Navier-Stokes equations for the viscous thin material layer\cite{jankuhn2018incompressible}:
\begin{equation} \label{momentum}
\left\{
\begin{split}
 \rho \dot \bu & =- \nabla_\Gamma  p + 2\mu \divG (\bD_\Gamma(\bu))  + \bb +   p \kappa\bn, \\
 \divG \bu  & =0,\\
 \dot{\rho} & =0,
\end{split}\right.\quad\text{on}~\Gamma(t).
\end{equation}
Here, $\kappa$ is the double mean curvature, $p$ is the surface pressure, $\mu$ is the viscosity, $\bD_\Gamma(\bu)=\frac12(\nabla_\Gamma \bu + \nabla_\Gamma \bu^T)$ is the surface rate-of-strain tensor \cite{GurtinMurdoch75}, {and $\bb$ represents area forces, which may include elastic and external forces}. If $\rho$ is smoothly extended to a space-time neighborhood of $\Gamma(t)$, then the following identity holds:
\[
\dot{\rho} = \frac{\partial \rho}{\partial t}+(\bu\cdot\nabla)\rho,
\]
which is independent of the particular extension. The same identity holds for the componentwise material derivative of $\bu$.

The geometric evolution of the surface is defined by the normal velocity $V_\Gamma$ of $\Gamma$, which is given by the normal component of the material velocity $\bu$:
\begin{equation} \label{Gamma_evol}
V_\Gamma=\bu\cdot \bn \quad\text{on}~\Gamma(t).
\end{equation}
If $\bb$ is given or defined through other unknowns, then equations \eqref{momentum}--\eqref{Gamma_evol} form a closed system of six equations for six unknowns: $\bu$, $p$, $\rho$, $V_\Gamma$, subject to suitable initial conditions. {Note that the domain $\Gamma(t)$ where the system is posed is not known {\it a priori} but is defined by $V_\Gamma$ and the initial surface $\Gamma(0)$. We emphasize this observation by counting $V_\Gamma$ as an independent unknown, although it is simply equal to the normal part of $\bu$. Despite its appearance, the system \eqref{momentum}--\eqref{Gamma_evol} is (strongly) nonlinear through the presence of the material derivative $\dot{\bu}$, the dependence of the surface $\Gamma(t)$ on $\bu$, and a possible dependence of $\bb$ on the shape of $\Gamma(t)$. We assume} that $\rho=\mbox{const}$ at $t=0$, and then $\dot{\rho} =0$ implies that the density stays constant for all times:
\[
\rho=\mbox{const}.
\]
{This assumption of uniform density is applied further throughout the paper.}

We further {distinguish} between area forces $\mathbf{b}$ coming from the adjacent inner--outer media and elastic forces generated by the bending and stretching of the membrane,
\begin{equation} \label{forces}
\bb=\bb^{\rm ext}+\bb^{\rm elst}.
\end{equation}
{Since the purpose of this paper is finding equilibrium configurations}, we assume the external force given by a constant pressure difference across the membrane,
	\begin{equation} \label{force1}
		\mathbf{b}^{\mathrm{ext}}= p^{\mathrm{ext}}\mathbf{n},\quad \text{with}~~p^{\mathrm{ext}}=\mathrm{const}.
	\end{equation}
	For the elasticity, we consider the Helfrich model with the Willmore energy functional~\cite{canham1970minimum,helfrich1973elastic}:
	\begin{equation} \label{Willmore}
		H=\frac{c_\kappa}{2}\int_{\Gamma}(\kappa-\kappa_0)^2, \mathrm{d}s+c_K\int_{\Gamma}K, \mathrm{d}s,
	\end{equation}
	where $K$ is the Gauss curvature, and material parameters $c_\kappa>0$, $c_K>0$, $\kappa_0$ have the meaning of bending rigidity, Gaussian bending rigidity, and spontaneous curvature, respectively. In this paper we restrict our interest to the simplest model with
\[
\kappa_0=0.
 \]
For closed surfaces that do not change their topology during evolution, the Gauss--Bonnet theorem implies that the second term in the Willmore energy functional equals $2\pi$ times the surface Euler characteristic. Hence this term does not contribute to the variation of the energy and thus to the elastic forces.
By the principle of virtual work, we obtain
\[
\int_{\Gamma(t)}\bb^{\rm elst}\cdot\bv\,ds= 
-\left.\frac{dH}{d\Gamma}\right|_{\bv},
\]
where $\frac{\mathrm{d}H}{\mathrm{d}\Gamma}|_{\mathbf{v}}$ is the variation of the energy functional on the (infinitesimal) displacement of $\Gamma$ given by the vector field $\mathbf{v}$.

The shape derivative of $H$ can be computed to take the form of
\begin{equation}\label{BE1}
\left.\frac{dH}{d\Gamma}\right|_{\bv} = c_\kappa\int_{\Gamma(t)}(-\Delta_\Gamma \kappa -\frac12 \kappa^3+2K\kappa)(\bv\cdot\bn)\,ds.
\end{equation}
The result in \eqref{BE1} is well-known due to Willmore\cite{willmore1996riemannian}. For completeness, we give in Appendix a short  proof using elementary tangential calculus.
From \eqref{BE1} it is clear that the release of the bending energy produces a force  in the normal direction to the surface:
\begin{equation} \label{bN}
	\bb^{\rm elst}=c_\kappa(\Delta_\Gamma \kappa +\frac12 \kappa^3-2K\kappa)\bn.
\end{equation}

\begin{rem}\rm The fluid system \eqref{momentum}--\eqref{Gamma_evol} with a generic force $\mathbf{b}$ was independently derived in Ref.\cite{jankuhn2018incompressible} from balance laws of continuum mechanics and in Ref.\cite{Gigaetal} from energetic principles (with $\mathbf{b}=0$). Equations for moving fluid membrane were also derived in local (curvilinear) coordinates~\cite{hu2007continuum,nitschke2019hydrodynamic}. If translated into the language of tangential calculus, the equations from Refs. \cite{hu2007continuum,nitschke2019hydrodynamic} were shown\cite{brandner2022derivations}  to also yield \eqref{momentum}--\eqref{Gamma_evol}. A relation between different formulations found in the literature was also discussed in Ref.\cite{reuther2018erratum}.
\end{rem}

\subsection{Conditions of equilibrium} \label{splitting}
We are interested in the equilibrium state solutions to \eqref{momentum}--\eqref{Gamma_evol} with external and bending forces \eqref{force1}, \eqref{bN}. The \emph{geometric} equilibrium requires $\Gamma(t)$ to be time-independent (in the sense of a shape). This and \eqref{Gamma_evol} immediately implies the  \emph{first equilibrium condition}:
\begin{equation}\label{cond1}
\bu\cdot\bn=0.
\end{equation}

\textbf{Case $\bu=0$.} Let us start with considering the case of no-flow, $\bu=0$. The momentum equation in \eqref{momentum} yields
$\nabla_\Gamma  p=	\bb^{\rm elst} +   (p^{\rm ext}+ p \kappa)\bn$. The left hand side of this identity is tangential to $\Gamma$, while the right hand side is orthogonal, and so both are zero yielding  $ p=\mbox{const}$ (the implication of $\nabla_\Gamma  p=0$ along $\Gamma$). Denote this constant surface pressure by $ p_0$. Then
$0=	\bb^{\rm elst} +   (p^{\rm ext}+ p \kappa)\bn$ and \eqref{bN} imply
\begin{equation}\label{equl1}
	c_\kappa(\Delta_\Gamma \kappa +\frac12 \kappa^3-2K\kappa) +  p_0 \kappa +p^{\rm ext}=0.
\end{equation}
Equation \eqref{equl1} was also derived in Ref.\cite{zhong1989bending} as the optimality condition for finding the minimum of Willmore energy \eqref{Willmore} subject to conserved surface area and enclosed volume, with constants $p_0$ and $p^{\rm ext}$ playing the role of Lagrange multipliers for the area and volume constraints, respectively. This constrained minimization problem has been extensively studied in the literature in the context of finding the shapes of vesicles\cite{jenkins1977static,luke1982method,peterson1985instability,seifert1991shape,seifert1997configurations}.

We conclude that for the static equilibrium ($\bu=0$) the system \eqref{momentum}--\eqref{Gamma_evol}, \eqref{force1}, \eqref{bN} coincides with a well-studied problem of Willmore energy constrained minimization. The membrane fluidity does not play a role in this scenario. We now consider the case of dynamic equilibrium.

\textbf{Case $\bu\neq0$.} The geometric equilibrium condition still implies $\bu\cdot\bn=0$ (only lateral motions are allowed).
To deduce other conditions, we  consider the tangential part of the momentum equation \eqref{momentum}.
This can be done by applying the orthogonal projection $\bP$ to the first equation in  \eqref{momentum} and noting that $\bP\bn=0$ and hence
$\bP\bb=0$. We get
\[
\rho \bP\dot \bu =- \nabla_\Gamma  p + 2\mu \bP\divG (\bD_\Gamma(\bu)).
\]
For tangential field $\bu$ we have $\bP\bu=\bu$  and so $\bP(\bu\nabla \bu)= \bP(\nabla \bu)\bu=\bP(\nabla \bu)\bP\bu =(\nabla_\Gamma \bu) \bu$.  {For a geometrically  stationary surface we also have} $\frac{\partial\bP}{\partial t}=0$. {This and  $\bP(\bu\nabla \bu)= (\nabla_\Gamma \bu) \bu$} imply the following identity for the projection of material derivative:
\[
	\bP\dot \bu  = \bP\Big(\frac{\partial \bu}{\partial t} + (\nabla \bu) \bu\Big) = \frac{\partial \bu}{\partial t} + (\nabla_\Gamma \bu) \bu = \frac{\partial \bu}{\partial t}+(\bu\cdot\nabla_\Gamma)\bu.
\]
We therefore get the following system satisfied by $\bu$, {such that $\bu\cdot\bn=0$,} and $ p$:
\begin{equation}\label{NSstat}
	\left\{
	\begin{aligned}
		\rho \left( \frac{\partial \bu}{\partial t}+(\bu\cdot\nabla_\Gamma)\bu \right)&= -\nabla_\Gamma  p + 2\mu\bP \divG \bD(\bu)\\
		\divG \bu  &= 0
	\end{aligned}\right.
\end{equation}
{on} a geometrically stationary $\Gamma$. {The system \eqref{NSstat} is the Navier--Stokes equations on a Riemann  manifold~\cite{chan2017formulation}.}

Multiplying the first equation in \eqref{NSstat} with $\bu$, integrating over $\Gamma$ and integrating by parts brings us to the energy equality
\[
\frac\rho2\frac{d}{dt}\int_{\Gamma} |\bu|^2\,ds=-2\mu \int_{\Gamma} |\bD_\Gamma(\bu)|^2ds.
\]
We see that the kinetic energy of the lateral flow decays for all motions with $\bD_\Gamma(\bu)\neq0$.
Therefore,  the equilibrium flow must satisfy  the  \emph{second equilibrium condition}:
\begin{equation}\label{Killing}
	\bD_\Gamma(\bu)=0.
\end{equation}
`Tangentially rigid' motions satisfying \eqref{cond1} and \eqref{Killing} correspond to Killing vector fields on manifolds~\cite{eisenhart1997riemannian,sakai1996riemannian}.
A non-zero Killing  field  generates a continuous one-parameter group of transformations $\Gamma\to\Gamma$ which are isometries, and  the ability of $\Gamma$ to support it is a \emph{geometric} constraint.
In particular,  among 2D compact closed surfaces only those of genus 0 and 1 may have non-zero  Killing fields and  the corresponding   group of transformations is 1 with the exception of surfaces of constant  curvature, i.e. those isometric to a sphere (3 parameter group) or a flat torus (2 parameter group)~\cite{myers1936isometries}. Moreover, the intrinsic geometry of such surfaces is rotationally symmetric, see e.g. Ref.\cite{eisenhart1997riemannian} and lemma~0.1 in Ref.\cite{chen2006note}. Additional assumptions on the Gauss curvature ensure (see Ref. \cite{nirenberg1953weyl} where the proof is given if $K>0$ on $\Gamma$ or more recent treatment in Ref.\cite{engman2004note})  that there is a unique smooth isometric embedding of such surface into $\mathbb{R}^3$ as a classical surface of revolution. We have not found results in the literature from which one may conclude that without additional assumptions on $K$ the classical surface of revolution is the only representation in $\R^3$ of a connected compact closed smooth surface with a Killing field, although such conclusion looks very plausible. We note that Killing fields often appear in the studies of fluid equations on manifolds\cite{jankuhn2018incompressible,reuther2018solving,olshanskii2019penalty,samavaki2020navier,pruss2021navier}.


With the help of $\gradG |\bu|^2 = 2 (\gradG \bu)^T\bu$, {which holds if $\bu\cdot\bn=0$,} and $(\bu\cdot\nabla_\Gamma)\bu= (\gradG \bu)\bu$ one verifies the identity
\[
(\bu\cdot\nabla_\Gamma)\bu= 2\bD_\Gamma(\bu)\bu-\frac12\gradG |\bu|^2.
\]
Using this identity in \eqref{NSstat} we see that for steady flow fields satisfying \eqref{Killing}  the momentum equation reduces to $\gradG  p-\frac\rho2\gradG |\bu|^2=0$. We thus get our \emph{third equilibrium condition}:
\begin{equation}\label{pressure}
	 p-\frac\rho2 |\bu|^2= p_0
	\qquad\text{with some}\quad p_0:=const.
\end{equation}
According to \eqref{pressure} the in-surface pressure {\emph{in an equilibrium state}} splits into a constant term and a term representing the kinetic energy density. For a pure fluid membrane ($c_\kappa=0$), $ p$ can be interpreted as  the surface tension coefficient, which is found to depend on the in-plane flow.

Summarizing, we obtain three conditions  for the velocity and pressure  of the fluid membrane in an equilibrium. These conditions \eqref{cond1}, \eqref{Killing}, and \eqref{pressure} are independent of an elasticity model and we use them below together with the particular elasticity model to derive the shape equation.

\subsection{Shape equations} The  Weingarten mapping (shape operator) $\bH:\Gamma\to\mathbb{R}^{3\times3}$ is given by $\bH=\nabla_\Gamma\bn$.
Note that $\bH=\bH^T$, $\bH\bn=0$. Eigenvectors of $\bH$ orthogonal to $\bn$ are the principle directions on $\Gamma$ and the corresponding eigenvectors are the curvatures $\kappa_1$ and $\kappa_2$. In particular, $\kappa=\kappa_1+\kappa_2:=\tr(\bH)$.
We also need the following identity  for the material derivative of $\bn$  (see eq.~(2.14) in Ref.\cite{jankuhn2018incompressible}):
\begin{equation} \label{aux442}
	\dot{\bn}=\bH \bu - \gradG (\bu\cdot\bn) .
\end{equation}
To deduce the shape equation, we first take the normal part of the momentum equation \eqref{momentum},
\begin{equation} \label{aux372}
\rho \bn\cdot\dot\bu  = 2 \mu \bn\cdot \divG \bD_\Gamma(\bu) + p \kappa + \bn\cdot \bb.
\end{equation}
The first term on the right-hand side vanishes due to \eqref{Killing}. For the normal projection of the material derivative we compute with the help of  $\bu\cdot\bn=0$ and \eqref{aux442}
\[
0=\dot{(\bn\cdot\bu)}= \bn\cdot\dot\bu+\bu\cdot\dot\bn=\bn\cdot\dot\bu+\bu^T\bH \bu.
\]
Substituting this and \eqref{pressure}, \eqref{force1}, \eqref{bN} in \eqref{aux372} gives the \emph{shape equation}
\begin{multline} \label{reaction2}
  -\rho  \bu^T\bH \bu-\frac\rho2\kappa|\bu|^2\\=  p_0  \kappa+ c_\kappa(\Delta_\Gamma \kappa +\frac12 \kappa^3-2K\kappa)+p^{\rm ext}
\end{multline}
with some $p_0=\mbox{const}$ {and tangential velocity $\bu$}. {At equilibrium}, the term $\rho \bu^T \bH \bu$ on the left-hand side can be interpreted as the normal component of a centrifugal force generated by the material flow along a curved trajectory. This interpretation becomes evident when we restrict to axisymmetric shapes below. Therefore, the shape equation {\eqref{reaction2}} represents the balance between the normal component of the centrifugal force, the effective membrane tension $(p_0+\frac{\rho}{2}|\bu|^2)\kappa$, the bending force, and the osmotic pressure $p^{\rm ext}$. In turn, the effective membrane tension can be split into the 'static' term $p_0\kappa$ and the 'dynamic' term $\frac{\rho}{2}|\bu|^2\kappa$.

Summarizing, the problem of finding dynamic equilibrium of a fluid--elastic membrane can be formulated as follows:
For the given density $\rho$, bending rigidity $c_\kappa$, osmotic pressure $p^{\rm ext}$,   and  surface area $A=\mbox{area}(\Gamma)$ find a shape $\Gamma$, tangential flow $\bu$ and  parameter $ p_0$ that solve \eqref{Killing} and \eqref{reaction2}.
Alternatively, one may ask to find $\Gamma$, $\bu$, $ p_0$, and  $p^{\rm ext}$ such that \eqref{Killing} and \eqref{reaction2} hold with given $\rho$,  $\kappa$, $A=\mbox{area}(\Gamma)$ \emph{and} $V=\mbox{vol}(\Gamma)$.
\smallskip

Any surface of revolution supports a non-zero Killing field. Moreover, it looks plausible that {only} surfaces of revolution support  non-zero Killing fields among closed compact smooth surfaces isometrically embedded in $\R^3$; see the discussion following \eqref{Killing}.
This motivates us to restrict further considerations to such surfaces.
 Without loss of generality, we let $Oz$ to be the axis of symmetry for $\Gamma$. Then tangential $\bu$ satisfying \eqref{Killing} is a field of rigid rotations given by
\begin{equation}\label{velocity}
\bu(\bx)=w\,\be_z\times\bx,\quad\bx\in\Gamma,\quad
\end{equation}
with the angular velocity $w\,\be_z$. {With the exception of a sphere,  functions $\mathbf{u}$ in \eqref{velocity} represents the entire family of Killing fields on $\Gamma$. Henceforth, we consider only $\bu$ given by \eqref{velocity}.} It holds
\[
|\bu(\bx)| = |w|\,r,\quad\text{with}~r=\mbox{dist}(\bx,Oz).
\]
For an axisymmetric surface, the first principle direction is tangential to the generating curve and the second one is the azimuthal direction and coincides with the direction of {$\bu$}. Since the principle directions are given by the eigenvectors of $\bH$, the later observation implies $\bu^T\bH \bu=\kappa_2 |\bu|^2$.
Now \eqref{Killing} yields the  \emph{shape equation for an axisymmetric surface}:
\begin{multline} \label{ShapeEq}
	-\rho\Big(\kappa_2+\frac\kappa2\Big)(w\,r)^2 \\=  p_0  \kappa+c_\kappa(\Delta_\Gamma \kappa +\frac12 \kappa^3-2K\kappa)+p^{\rm ext},
\end{multline}
with some $ p_0=\mbox{const}$.
Thus, further in the paper we are interested in the following problem:  \emph{Find an axisymmetric $\Gamma$, $ p_0$, and  $p^{\rm ext}$ such that \eqref{ShapeEq} holds with given $\rho$,  $\kappa$, $|w|$, $A=\mbox{\rm area}(\Gamma)$ \emph{and} $V=\mbox{\rm vol}(\Gamma)$.}
We  remark that instead of prescribing $w$ one may consider the prescribed angular momentum (a conserved quantity).
In such formulation, $w$ should be treated as unknown.

\begin{rem}\rm For $\bu=0$ eqs. \eqref{reaction2} and \eqref{ShapeEq} naturally simplifies to \eqref{equl1}, which is the  optimality condition for  constrained minimization of the energy functional \eqref{Willmore} with conserved surface area and enclosed volume. However, for the general case of $\bu\neq0$, it is not clear how the shape equation can be related to an energy minimization problem. {A recent work by Krause et al.~\cite{krause2022numerical} deduced a variant of \eqref{ShapeEq} by recognizing Killing fields in \eqref{velocity} as equilibrium solutions of the surface Navier-Stokes equations on axisymmetric surfaces. The shape equation in Ref.\cite{krause2022numerical}, however, uses a generic surface pressure variable ($p_0$ and dynamic pressure do not appear). This makes the problem much harder to address numerically or relate to the classical constrained minimization problem for the ceasing lateral flow.
	}
\end{rem}
	
\begin{rem}[scaling] \rm
A scaling property well-known for \eqref{equl1} extends to  \eqref{ShapeEq}: If a triple $\{\Gamma,  p_0, p_{\rm ext}\}$ solves   \eqref{ShapeEq}, then  for any $R>0$ the triple $\{R^{-1}\Gamma, R^2 p_0, R^3p_{\rm ext}\}$  solves   \eqref{ShapeEq} with
$ w\to R^2 w$. 
Choosing  a representative solution with $\mbox{area}(\Gamma)=4\pi^2$, it is therefore convenient to parameterize solutions by their reduced volume
	\begin{equation}\label{Vref}
		\widehat V=\widehat V(\Gamma):= 3\mbox{vol}(\Gamma)/(4\pi),\quad  	\widehat V(\Gamma)\in (0,1],
	\end{equation}
	where $\widehat V=1$  corresponds to the unit sphere, a trivial solution of \eqref{ShapeEq}  for $ w=0$ and $ p_0$, $p^{\rm ext}$ satisfying $2 p_0+ Rp^{\rm ext}=0$. Same scaling argument holds for solutions of \eqref{Killing}, \eqref{reaction2}.
\end{rem}

\section{Parametrization of the shape equation and a numerical solve}\label{sec:num}   
An axisymmetric $\Gamma$ can be described by its profile curve
\[
s\to (r(s), z(s)),
\]
so that $\Gamma$ is generated by rotating the profile curve around the $z$-axis in $\mathbb{R}^3$. Assuming $s$ is the arc-length parameter, one computes (cf. Section 3C in Ref.\cite{kuhnel2015differential}) principle curvatures to be $\kappa_1=-r_{ss}z_s+r_{s}z_{ss}$, $\kappa_2=\frac{z_s}r$.
It is convenient to introduce the tilt angle $\psi(s)$ (an angle between the $Or$-axis and tangent vector to the profile curve). Writing geometric quantities in terms of $\psi$, one gets
\[
r_s=\cos\psi,\quad z_s=\sin\psi,\qquad
\kappa_1=\psi_s,\quad \kappa_2=\tfrac{\sin\psi}{r}.
\]
One also computes $\Delta_\Gamma \kappa=\frac1r(r\kappa_s)_s$.
Denote the length of the profile  curve by $L$. Then the  boundary conditions at $s=0$ and $s=L$ are obviously
$
r(0)=0$, $r(L)=0$, $\psi(0)=0$, $\psi(L)=\pi.
$
The area and volume of the surface  $\Gamma$ can be computed as $2\pi \int_{0}^L r\,ds$ and $\pi \int_{0}^L r^2\sin\psi\,ds$, respectively.
Now, we can formulate the problem of finding a stationary shape as follows:\\
   Given  an angular velocity $ w\ge0$,  surface area $A>0$ and  volume $V>0$ (satisfying the isoperimetric inequality $V\le 1/(6\pi^2)A^{\frac32}$, i.e. necessary condition for a surface to exist),  find $L\in\R_+$, ~$\psi(s), r(s) :[0,L]\to\R$, ~ $ p_0,p^{\rm ext}\in\R$  satisfying
   the following system of ODEs, integral and boundary conditions:
   \begin{align}
   	 -\rho w^2r&(\tfrac{1}2r\psi_s+\tfrac32\sin\psi) \notag\\
     =& p_0  \kappa+c_\kappa \big(r^{-1}(r\kappa_s)_s +\tfrac12\kappa^3-2K\kappa\big)+p^{\rm ext}, \label{eq1}\\ 
   	 r_s=&\cos\psi, \label{eq2}\\
   	 2\pi\int_{0}^{L}& r\,ds= A,\quad
   	 \pi \int_{0}^L r^2\sin\psi\,ds=V\label{eq4}\\
   	 r(0)=&\,0,\quad r(L)=0,\quad\psi(0)=0,\quad\psi(L)=\pi.	\label{eq5}
   \end{align}
with $\kappa=(\psi_s+\frac{\sin\psi}{r}),~ K=\frac{\psi_s\sin\psi}{r}$.

\begin{figure}
 	\vskip-1ex 		
 	\includegraphics[width=0.4\textwidth]{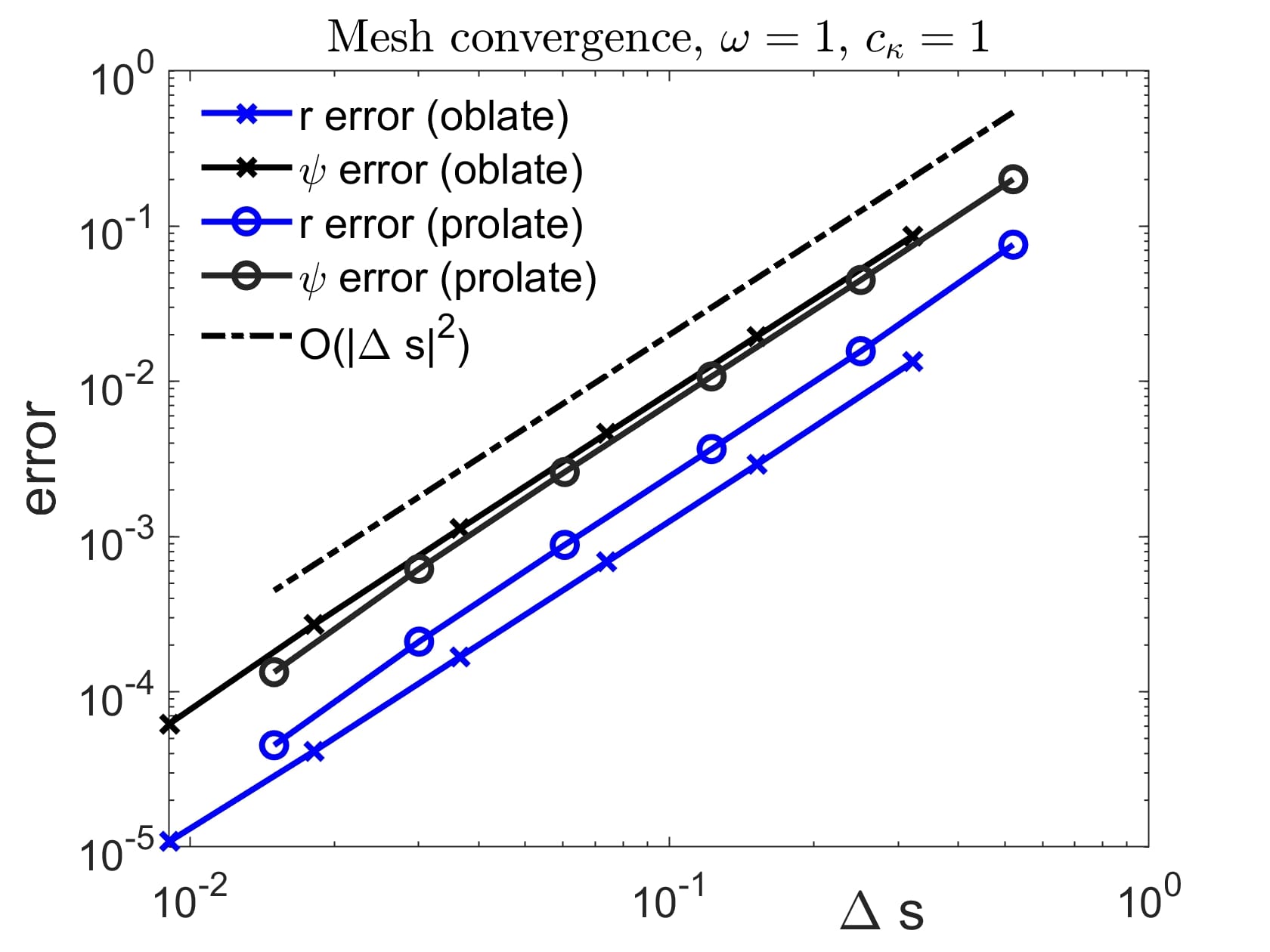}
 	\vskip-2ex 		
 	\caption{\label{fig1} Convergence of the numerical solutions for refined meshes.}
 	\vskip-1ex 		
 \end{figure}

The system \eqref{eq1}--\eqref{eq5} is further discretized using a staggered grid for $\psi$ and $r$ with a uniform
mesh step $\Delta s=L/N$.
We prescribed $r$-unknowns to nodes $x_i=i\Delta s$, {\small $i=0,\dots,N$} and $\psi$-unknowns 
 to nodes $\hat x_j=(j-\tfrac12)\Delta s$, {\small $j=0,\dots,N+1$}. Then equations  \eqref{eq1}--\eqref{eq2} are discretized (using standard finite differences)
in the inner $\psi$-nodes, integrals \eqref{eq4} are computed with the help of composite trapezoid and rectangular (using averaging for $r$ unknowns), respectively. After we approximate
boundary conditions in \eqref{eq5} by $r(x_0)=r(x_N)=0$, $\psi(\hat x_0)+\psi(\hat x_1)=0$, and
$\psi(\hat x_N)+\psi(\hat x_{N+1})=2\pi$, we obtain a non-linear system of $2N+6$ algebraic equations for $2N+6$ unknowns: $L$, $ p_0$, $p^{\rm ext}$,
$r(x_i)$, {\small $i=0,\dots,N$}, and $\psi(\hat x_j)$, {\small $j=0,\dots,N+1$}.  The system of algebraic equations is solved using a non-linear least-square method with the trust-region-dogleg algorithm, which finds search directions and is implemented in the `fsolve()' Matlab\texttrademark~ procedure. To verify the convergence of the numerical method, solutions were computed for a sequence of refined meshes with $N$ taking values from ${40,80,160,320,640}$. {The finest grid solution was taken as the reference, and the error was computed as the $\ell_\infty$ norm of the difference between} the solutions for $N\in{40,80,160,320}$ and the finest grid solution. The method demonstrates second-order convergence, as shown in Fig.~\ref{fig1}, for two examples of shapes, prolate and oblate.

 Assuming axial symmetry is a common approach to simplify the numerical study of minimal energy shapes. In particular, a shape parametrization  using $\psi$ and $r$ was employed in, e.g., Refs.\cite{luke1982method,peterson1985instability,miao1991equilibrium,seifert1991shape,julicher1996shape}. However, we believe that the numerical scheme presented in this work is novel.

\section{Stationary shapes}\label{sec:shapes}
To minimize the number of parameters we let
\[
c_\kappa\in\{0,1\},\quad \rho/2=1, \quad A=4\pi^2.
\]
This can be always ensured by a proper re-scaling of $ w$, $p_0$, and $p^{\rm ext}$. We then vary $ w$ and $\widehat{V}$ and solve \eqref{ShapeEq} to find $\Gamma$, $ p_0$, $p^{\rm ext}$.

\begin{figure}[h]
\begin{center}
	\includegraphics[width=0.43\textwidth]{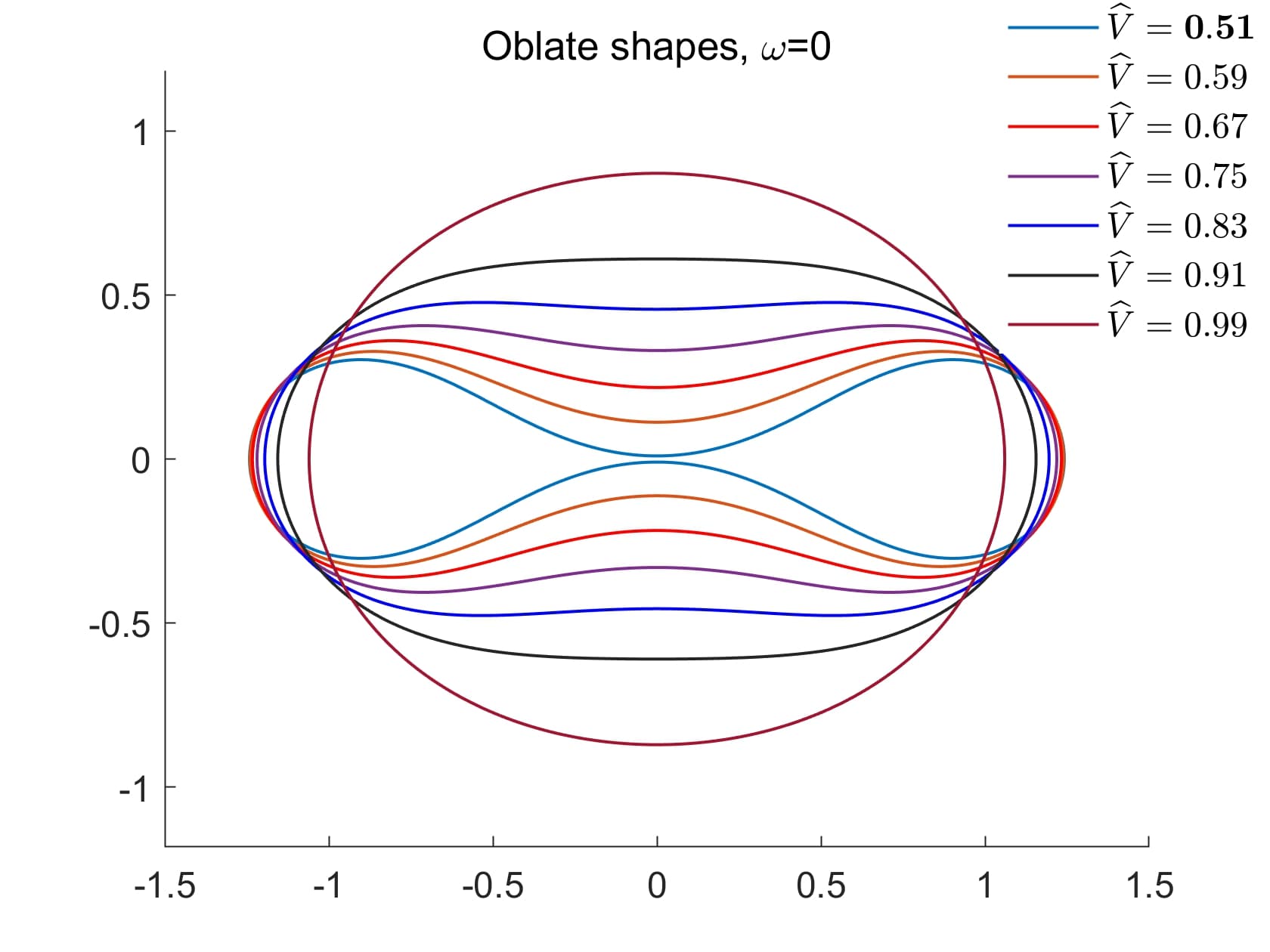}
	\includegraphics[width=0.45\textwidth]{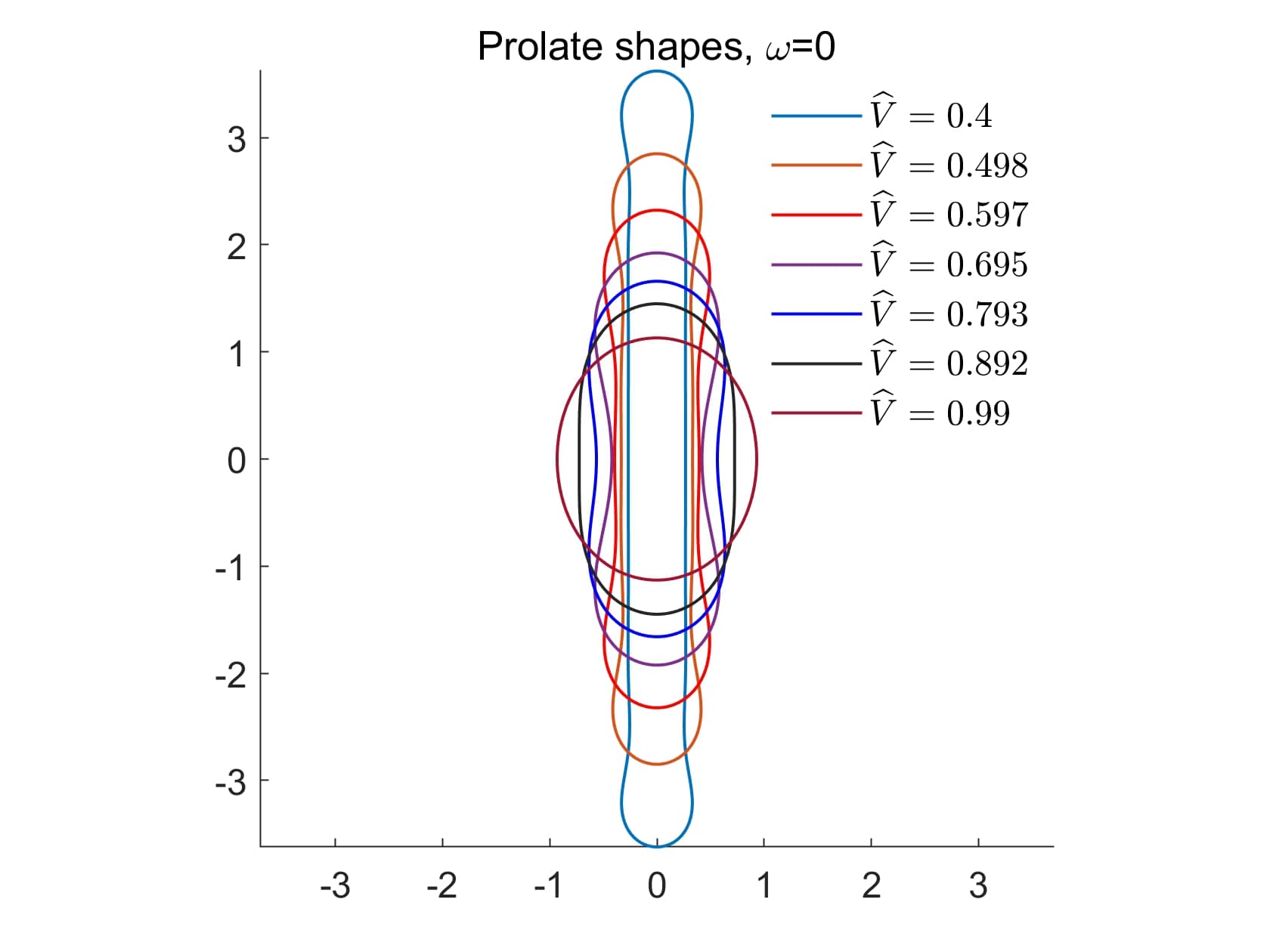}
\end{center}
\vskip-3ex	
	\caption{\label{fig2} Branches of oblate--biconcave and prolate--dumbbell shapes for pure elastic membrane.}
\end{figure}

\subsection*{Case $w=0$, $c_\kappa=1$.}
Setting $w=0$ (pure elasticity, no fluidity) results in two branches of solutions to \eqref{equl1}, consisting of oblate and prolate shapes, as shown in Fig~\ref{fig2}. To initiate each branch, we perturb the unit sphere by the second spherical harmonic as an initial guess for our nonlinear solver. The branch of oblate shapes continues with biconcave discocytes until approximately $\hat V\simeq 0.51$, while the branch of prolate shapes continues with increasingly elongated dumbbell forms. The resulting shapes and corresponding $p_0$ and $p^{\rm ext}$ are in perfect agreement with results known in the literature~\cite{jenkins1977static,seifert1991shape,seifert1997configurations}.

\subsection*{Case $w=4$, $c_\kappa=1$.}

\begin{figure}
	\begin{center}
		\includegraphics[width=0.4\textwidth]{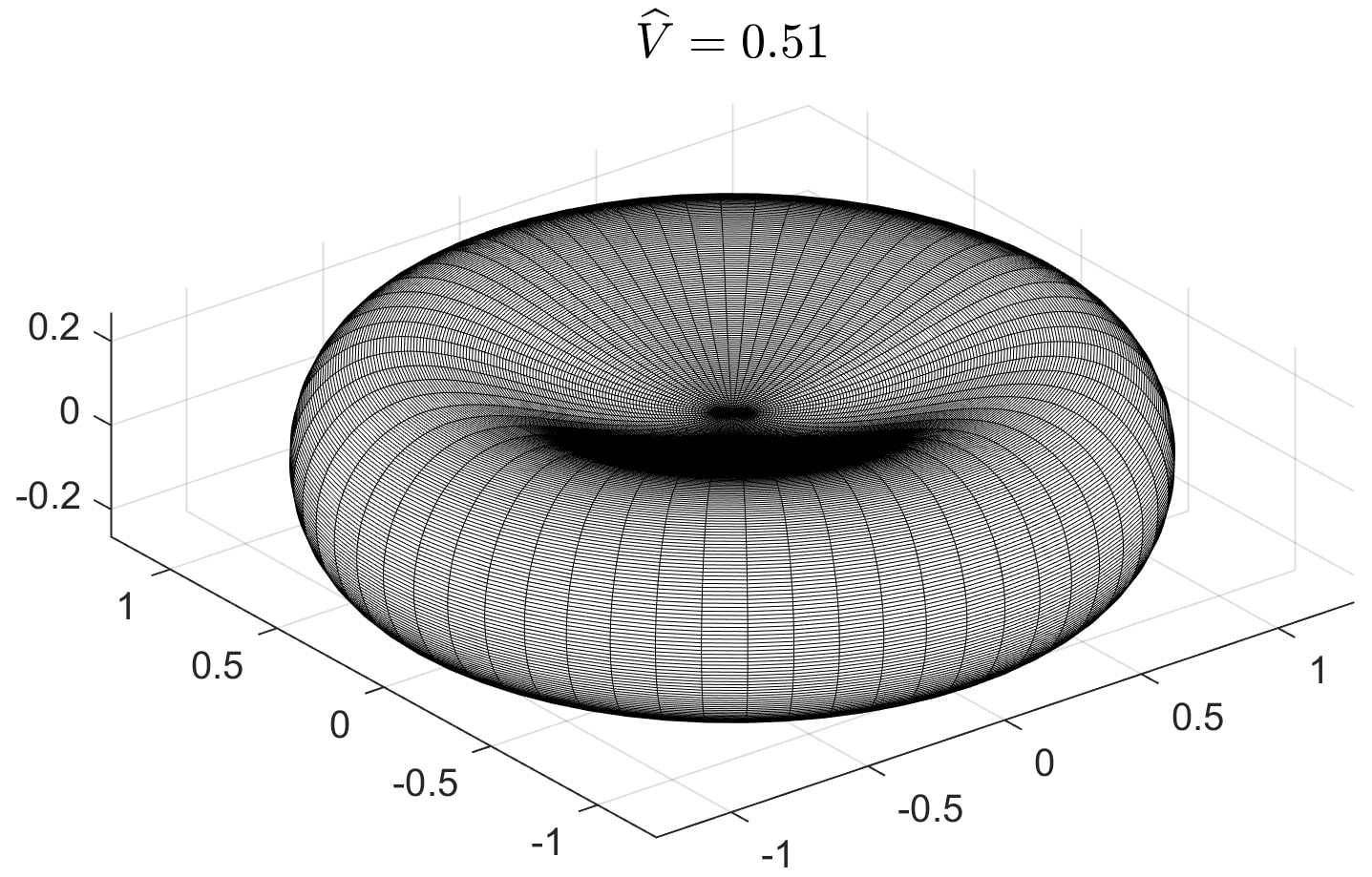}
		\includegraphics[width=0.45\textwidth]{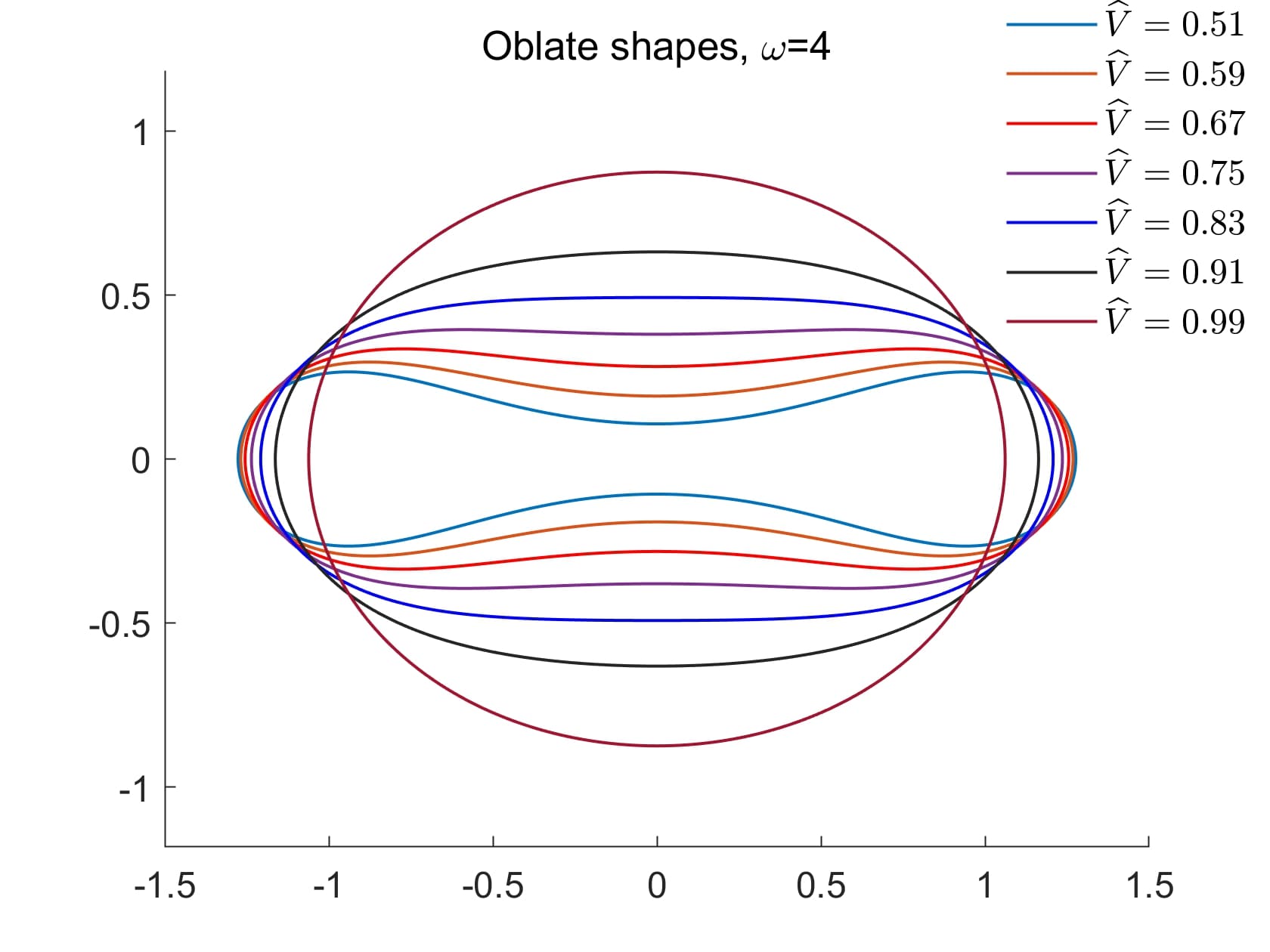}
	\end{center}
\vskip-3ex
	\caption{\label{fig3}\small  A branch of oblate--biconcave shapes for fluid--elastic membrane with $ w=4$. Top panel visualizes the 3D shape for  $\widehat V=0.51$.}
\end{figure}

\begin{figure*}
	\begin{center}
		\begin{minipage}[c]{0.7\textwidth}
			\includegraphics[width=0.31\textwidth]{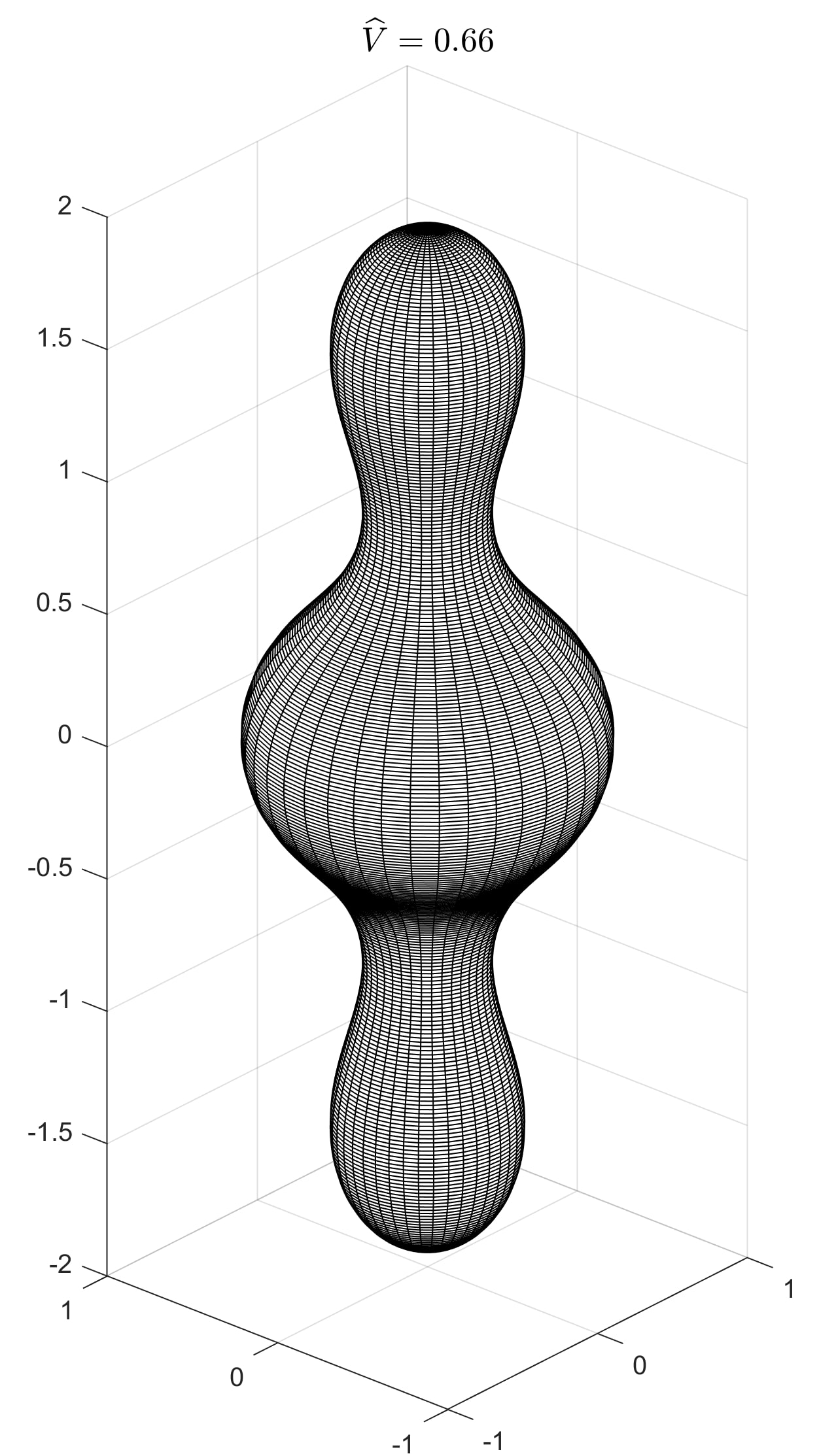}	\hskip-2ex	
			\includegraphics[width=0.69\textwidth]{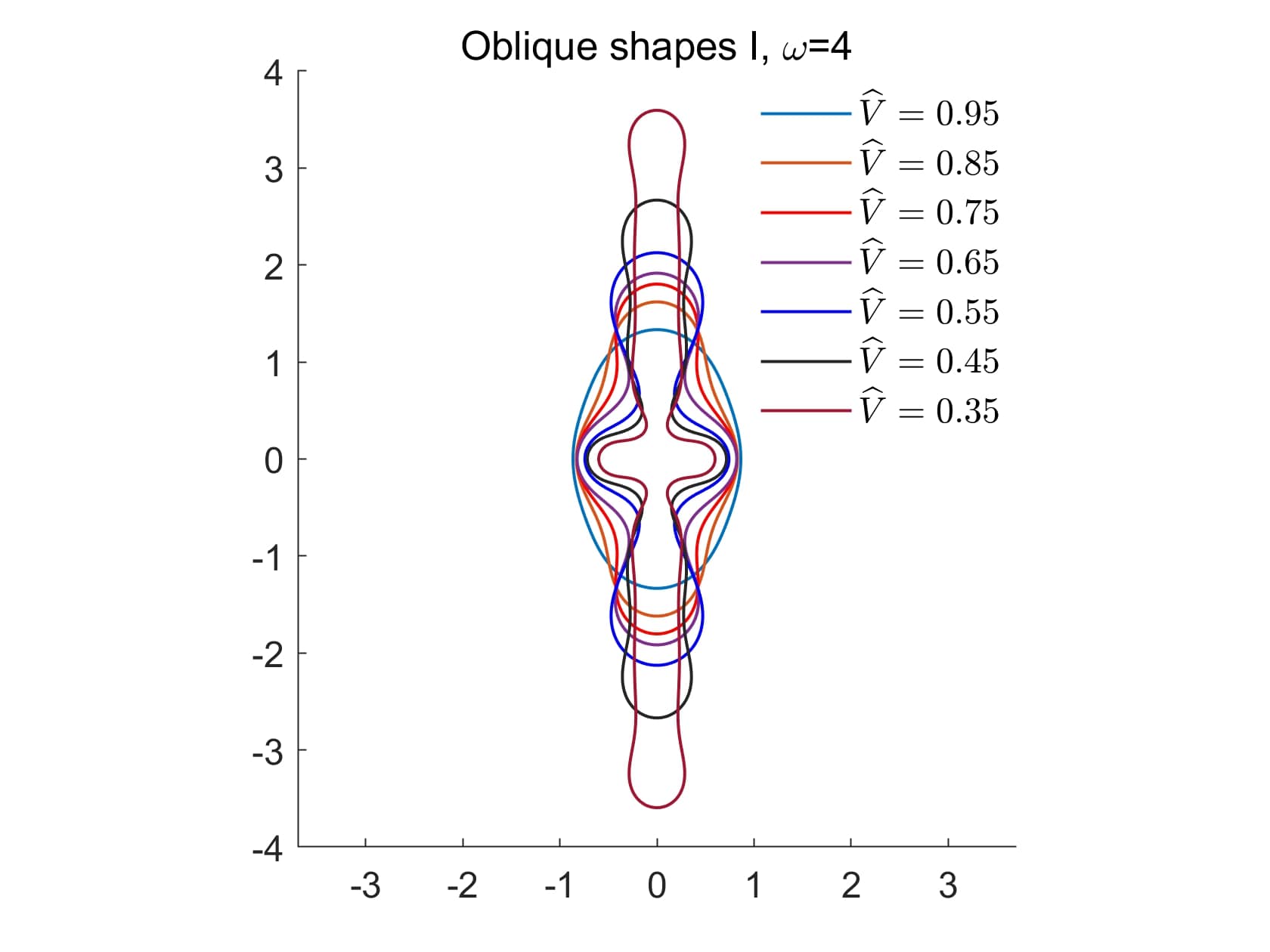}
		\end{minipage}\hskip-3ex
		\begin{minipage}[c]{0.2\textwidth}		\small
			\begin{tabular}{l|cc}	
				$\widehat{V}$& 	$ p_0$&	$p^{\rm ext}$\\ \hline
				0.35&   -174 & 29.0 \\
				0.45&  -95.7 & 18.9 \\
				0.55&  -74.8 & 17.0  \\
				0.65&  -39.9  & 8.25  \\
				0.75&   -38.6 & 9.11 \\
				0.85&   -42.6 & 11.8 \\
				0.95&   -56.1 & 18.6 \\
			\end{tabular}
		\end{minipage}
	\end{center}
\vskip-3ex
	\caption{\label{fig8}\small The first branch of oblique shapes for $ w=4$.  Left panel visualizes the 3D shape with  $\widehat V=0.66$.}
\end{figure*}

We now set $w=4$ to investigate how the equilibrium state is affected by the balance between bending forces and forces generated by fluid motion. Starting from an oblate perturbation of the unit sphere, we find a branch of oblate ellipsoids that continues with biconcave shapes as the reduced volume $\widehat{V}$ decreases; see Fig.~\ref{fig3}. However, the transition to biconcave forms occurs later than for $w=0$. The surface begins to self-intersect for $\widehat{V}\lesssim0.41$. Similar oblate ellipsoidal shapes were reported in Ref.\cite{krause2022numerical} as limit equilibrium solutions to the full system \eqref{momentum}--\eqref{Gamma_evol} with $\bb^{\rm ext}=0$ and $\bb^{\rm elst}$ as in \eqref{bN}. These solutions were obtained as stationary limits of 3D numerical solutions that start from a spherical shape with a Killing field as the initial condition.

\begin{figure*}
	\begin{center}
		\hskip-2ex\begin{minipage}[c]{0.7\textwidth}		
			\includegraphics[width=0.34\textwidth]{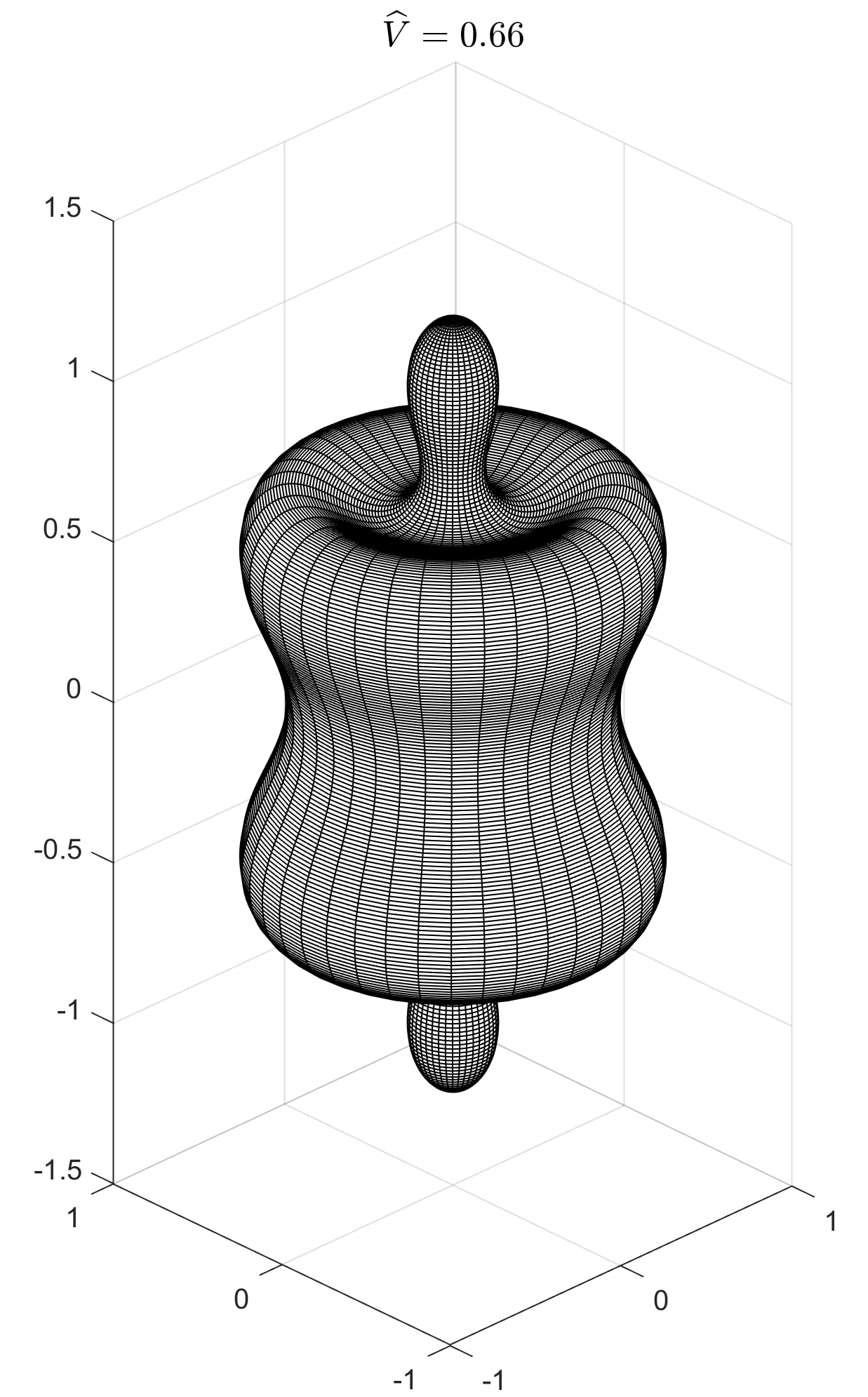}
			\includegraphics[width=0.61\textwidth]{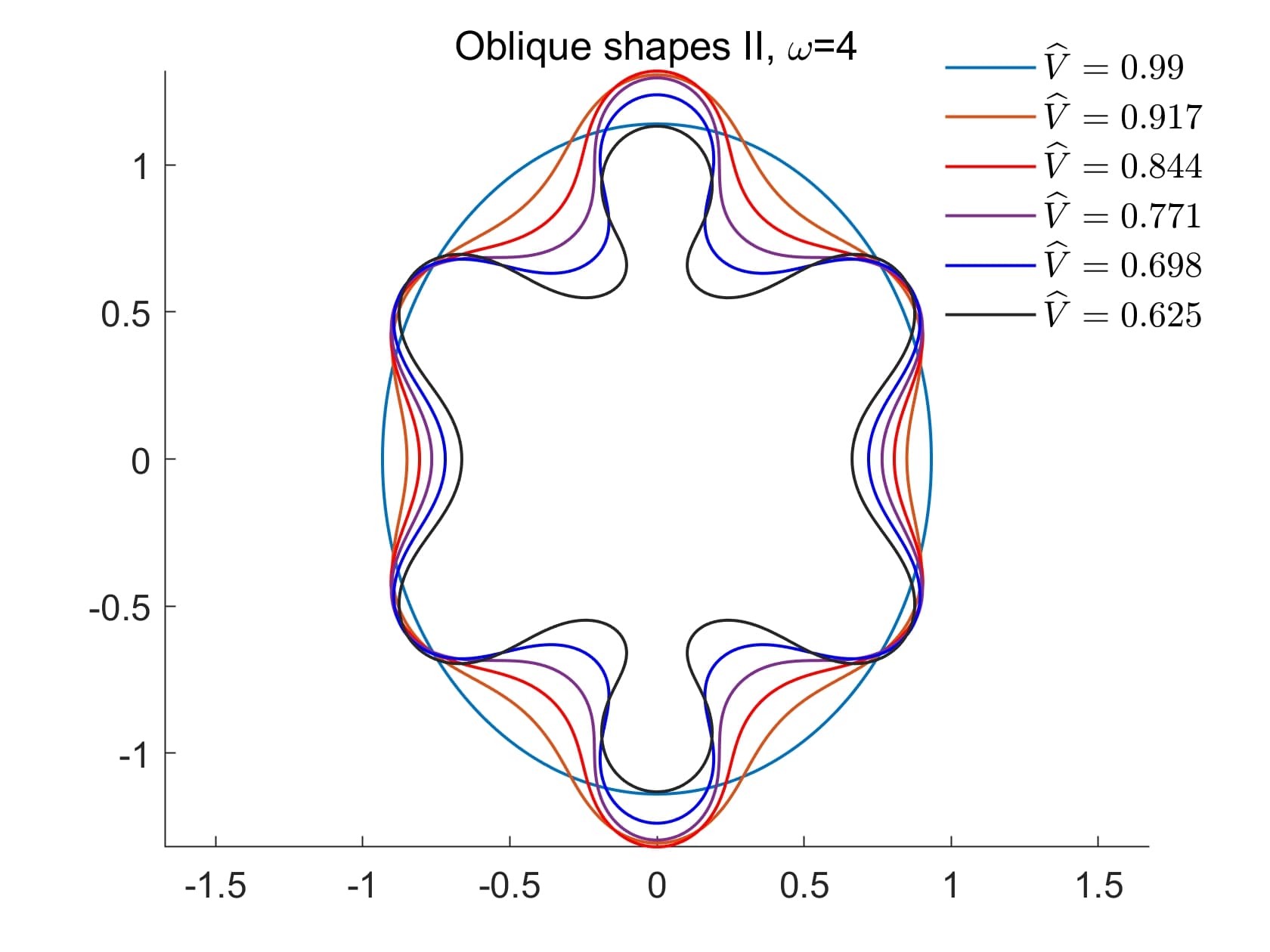}
		\end{minipage}\hskip-1ex
		\begin{minipage}[c]{0.2\textwidth}		\small
			\begin{tabular}{l|cc}	
				$\widehat{V}$& 	$ p_0$&	$p^{\rm ext}$\\ \hline
				0.625&  72.2 & -198 \\
				0.698&  89.7 & -229 \\
				0.771&  102 & -260  \\
				0.844&  115 & -294  \\
				0.917&  130 & -330 \\
				0.99&   10.3 & -57.5 \\
			\end{tabular}
		\end{minipage}
	\end{center}
\vskip-3ex
	\caption{\label{fig9}\small The second branch of oblique shapes for $ w=4$.  Left panel visualizes the 3D shape with  $\widehat V=0.66$.}
\end{figure*}

Starting with a prolate perturbation of the sphere, we were unable to find a branch of prolate ellipsoidal shapes in the vicinity of the unit sphere. Instead, we discovered two branches of oblique forms, as shown in Figs.\ref{fig8} and~\ref{fig9}. We were unable to compute shapes on these branches much beyond the smallest reported reduced volumes, i.e., $\widehat {V}=0.35$ and $\widehat {V}=0.625$, respectively. It is worth noting that the limit shape is close to pearling, a phenomenon known for pure Helfrich membranes with nonzero spontaneous curvature~\cite{seifert1991shape}.

\begin{figure*}
	\begin{center}
		\hskip-1ex\begin{minipage}[c]{0.7\textwidth}		
			\includegraphics[width=0.34\textwidth]{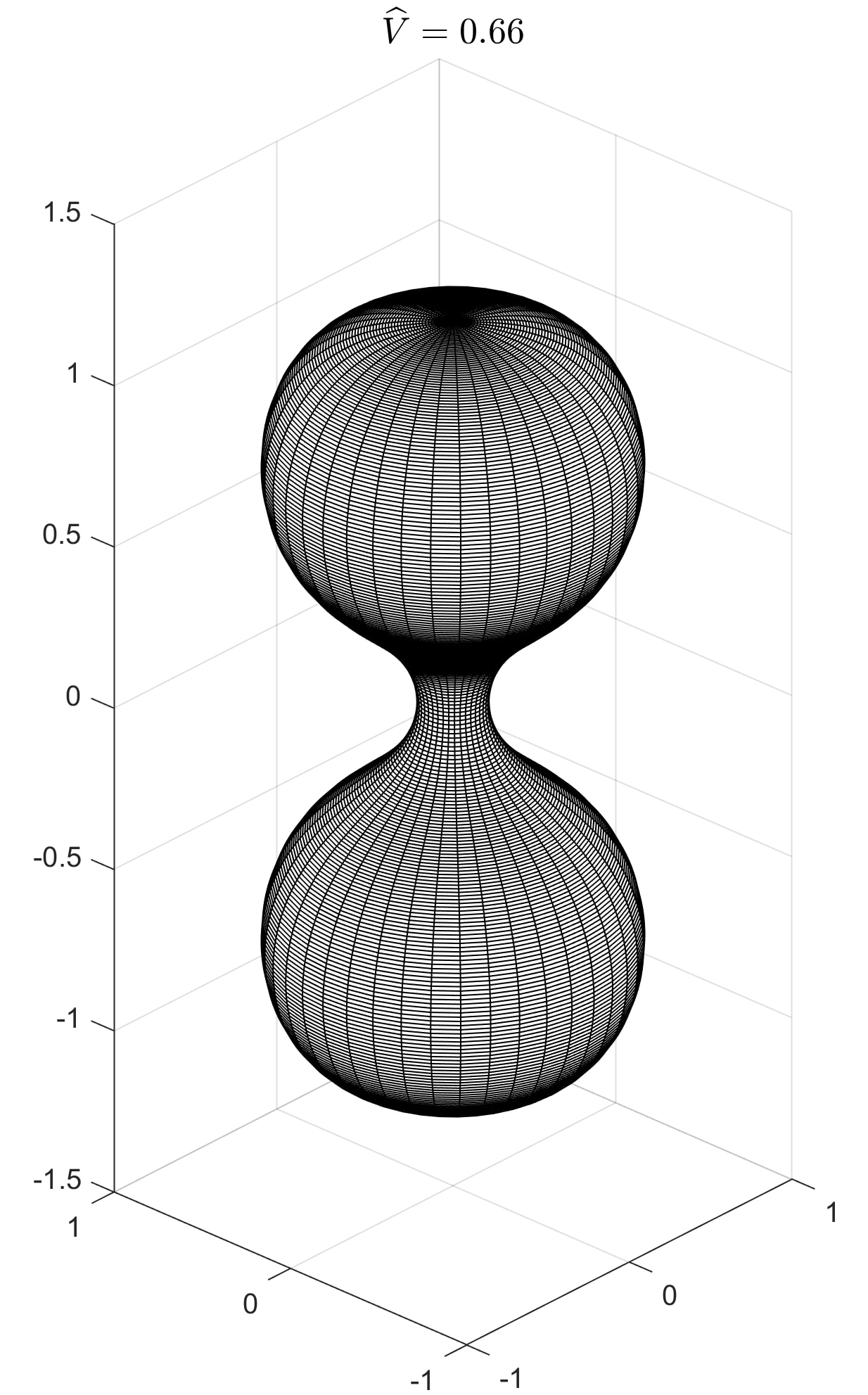}
			\includegraphics[width=0.61\textwidth]{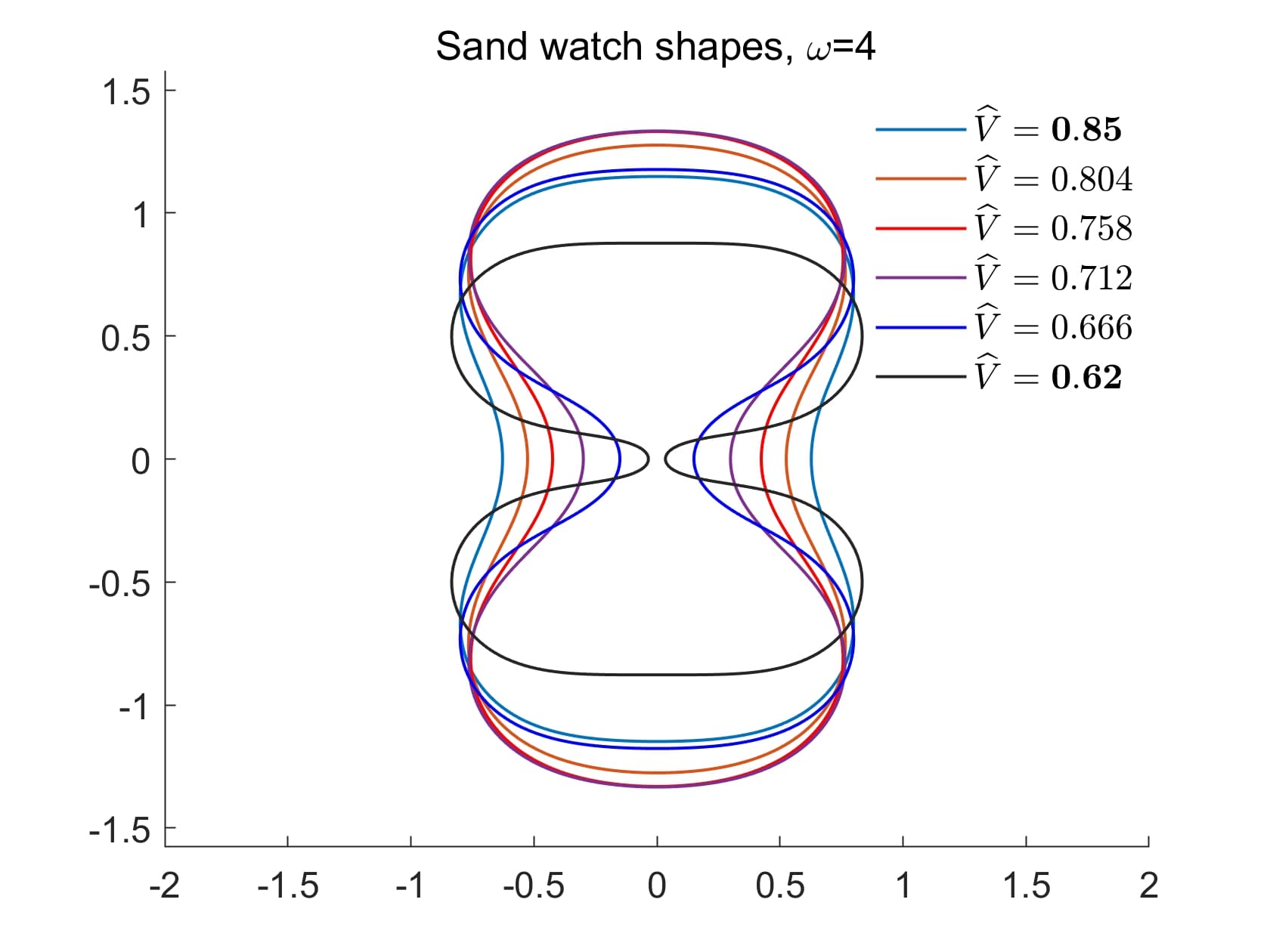}
		\end{minipage}\hskip-2ex
		\begin{minipage}[c]{0.2\textwidth}		\small
			\begin{tabular}{l|cc}	
				$\widehat{V}$& 	$ p_0$&	$p^{\rm ext}$\\ \hline
				0.62&   2.88 & -31.5 \\
				0.666&  3.25 & -28.8 \\
				0.712&  3.30 & -26.0  \\
				0.758&  4.34 & -27.5  \\
				0.804&  5.73 & -30.5 \\
				0.85&   7.82 & -35.9 \\
			\end{tabular}
		\end{minipage}
	\end{center}
\vskip-3ex
	\caption{\label{fig7}\small  A branch of sand watch shapes for $ w=4$.  Left panel visualizes the 3D shape with  $\widehat V=0.66$.}
\end{figure*}

Another branch of solutions was found for reduced volumes $\widehat {V}\in[0.62,0.85]$. The branch consists of sand watch shapes, as shown in Fig.~\ref{fig7}. For $\widehat {V}<0.62$ the neck of the shape is closing.
The non-linear solver also failed to converge to any solution for reduced volumes  larger than  $\widehat {V}\approx 0.85$.

\begin{figure*}
	\begin{center}
\begin{minipage}[c]{0.68\textwidth}		
		\includegraphics[width=0.18\textwidth]{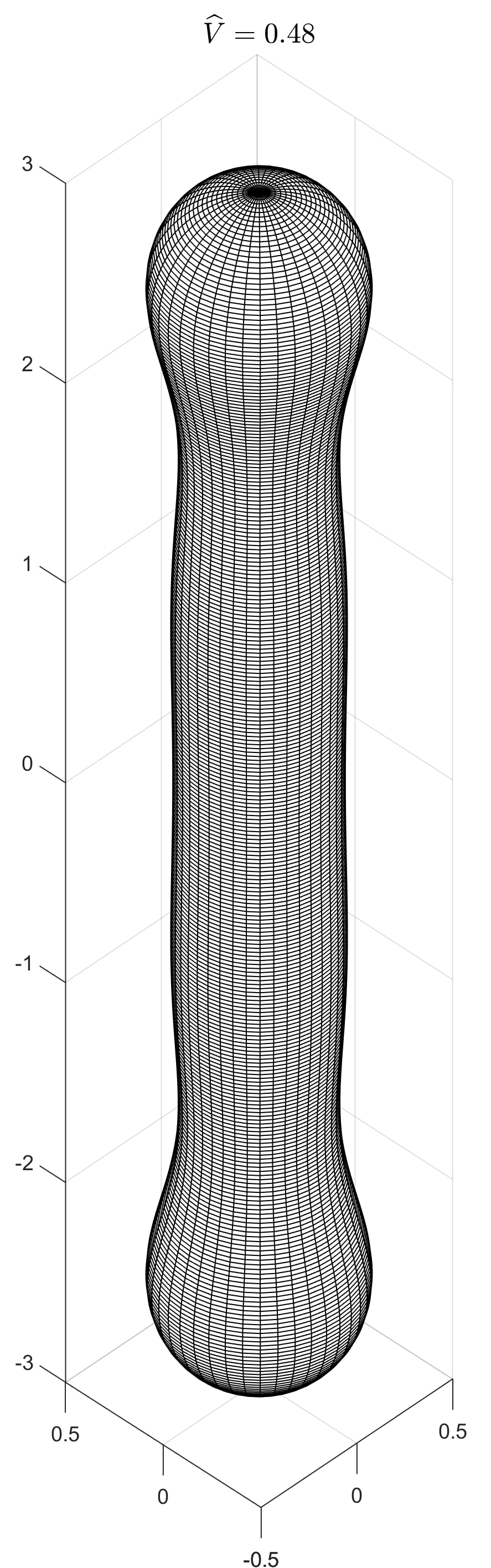}
		\includegraphics[width=0.75\textwidth]{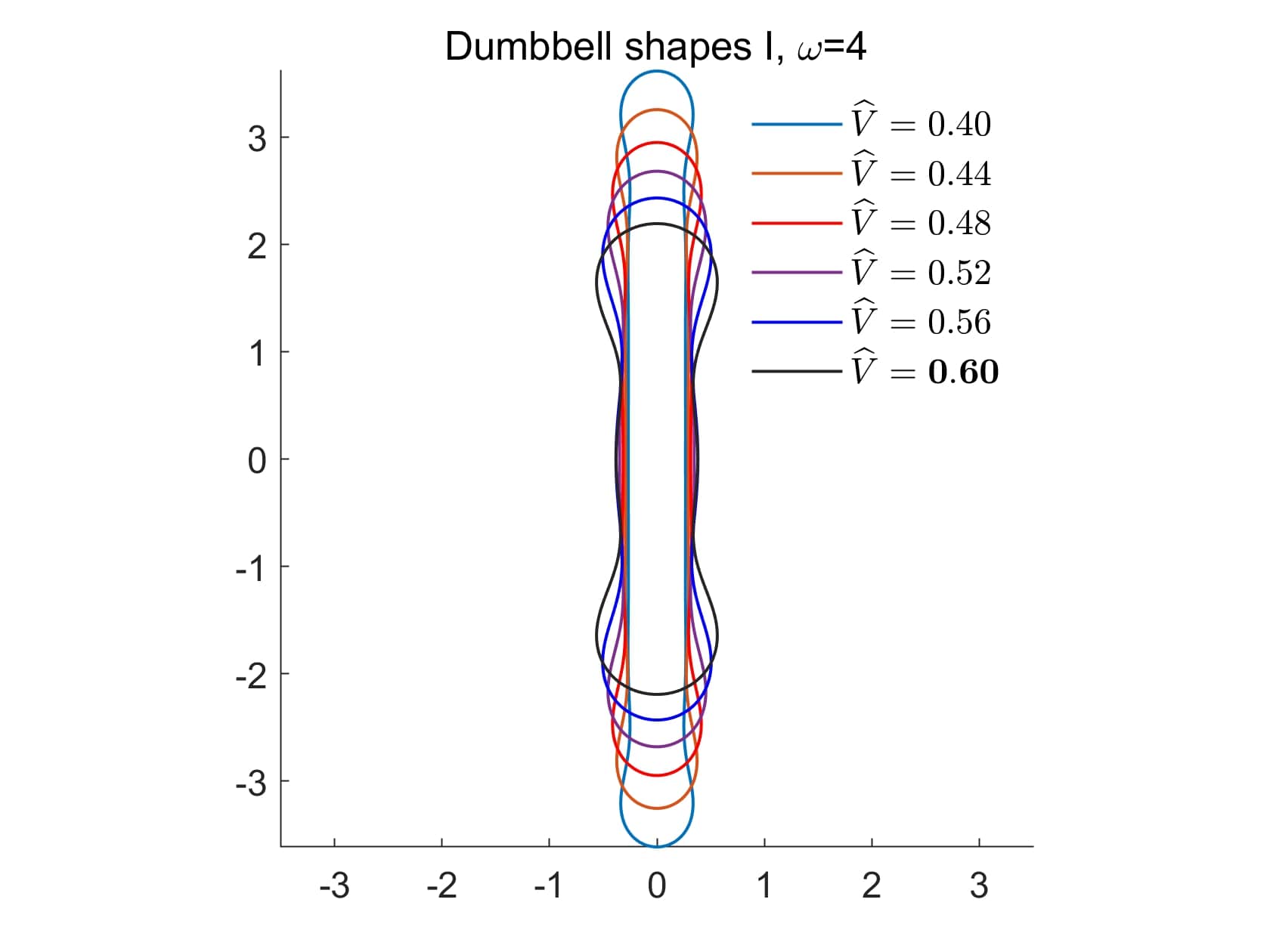}
\end{minipage}\hskip-5ex
\begin{minipage}[c]{0.2\textwidth}		\small
\begin{tabular}{l|cc}	
	$\widehat{V}$& 	$ p_0$&	$p^{\rm ext}$\\ \hline
	0.40&  21.8 & -115 \\
	0.44&  18.3 & -89.8 \\
	0.48&  15.7 & -72.9  \\
	0.52&  13.8 & -61.4  \\
	0.56&  12.5 & -53.7 \\
	0.60&  11.4 & -48.3 \\
\end{tabular}
\end{minipage}
	\end{center}
\vskip-3ex
	\caption{\label{fig5}\small  The first branch of dumbbell shapes for $ w=4$. Left panel visualizes the 3D shape with $\widehat V=0.48$.}
\end{figure*}

\begin{figure*}
	\begin{center}
	\begin{minipage}[c]{0.7\textwidth}
	\includegraphics[width=0.25\textwidth]{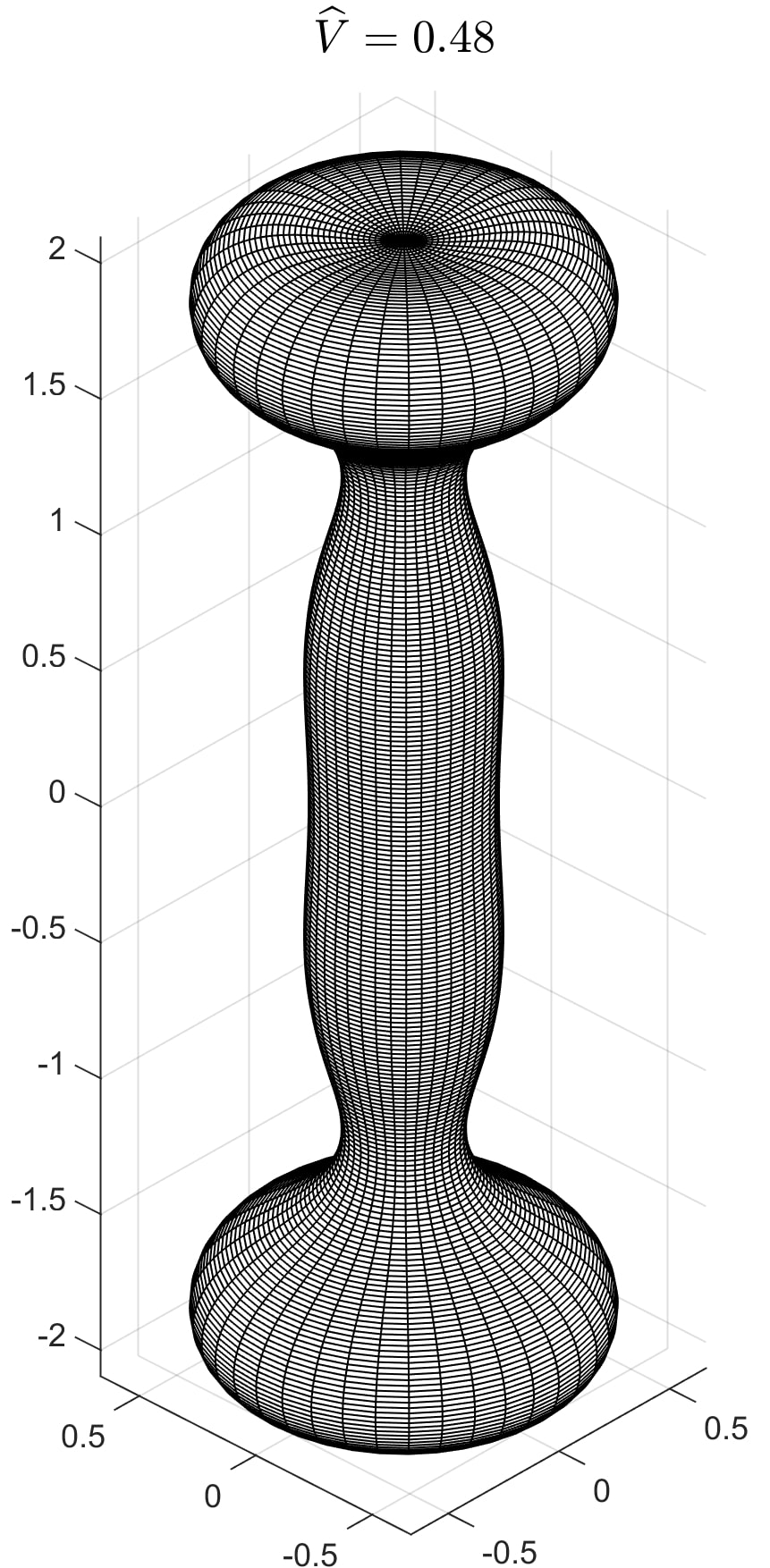}
	\includegraphics[width=0.7\textwidth]{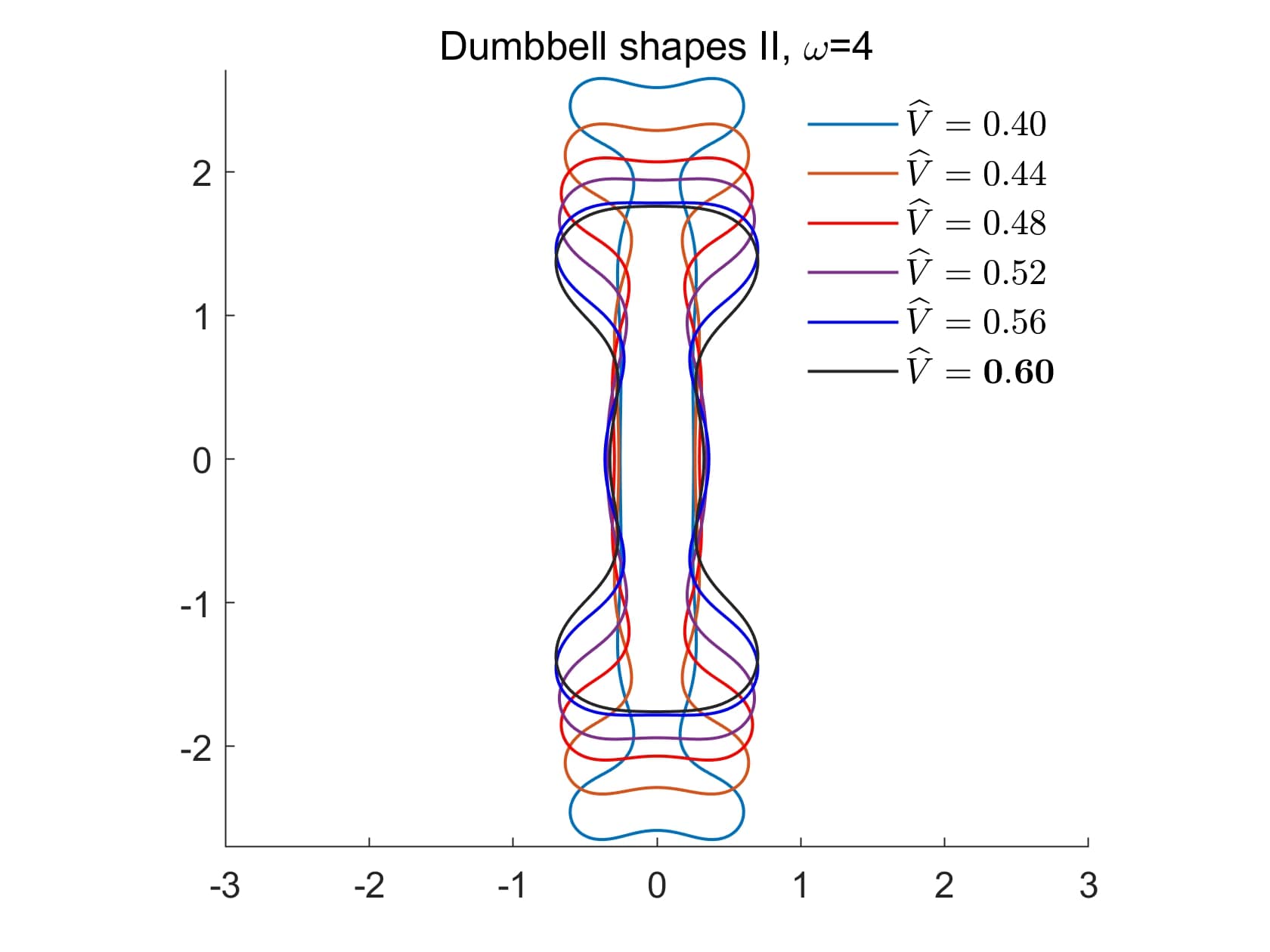}
	\end{minipage}\hskip-5ex
\begin{minipage}[c]{0.2\textwidth}		\small
	\begin{tabular}{l|cc}	
		$\widehat{V}$& 	$ p_0$&	$p^{\rm ext}$\\ \hline
		0.40&  12.8 & -57.2 \\
		0.44&  14.0 & -64.6 \\
		0.48&  16.7 & -78.2  \\
		0.52&  18.2 & -89.5  \\
		0.56&  20.7 & -107 \\
		0.60&  23.8 & -130 \\
	\end{tabular}
\end{minipage}
\vskip-3ex
	\end{center}
	\caption{\label{fig6}\small  The second branch of dumbbell shapes  at $ w=4$.  Left panel visualizes the 3D shape with $\widehat V=0.48$.}
\end{figure*}

Two branches of dumbbell shapes were found for reduced volumes $\widehat V\leq 0.6$, as shown in Figs.\ref{fig5} and~\ref{fig6}. The first branch somewhat resembles the dumbbell shapes found for $w=0$; compare the shape profiles in Fig.\ref{fig5} and Fig.\ref{fig2}. In the second branch, the dumbbell surfaces are distinctly different, featuring flatter concave discs as the reduced volume decreases. Our results suggest that around $\widehat V\simeq0.61$, there may be transition points where the oblique II and sand watch shapes yield two branches of dumbbell shape solutions.

In the present study, we do not address an important question regarding the stability of the newly found equilibrium states. It should be noted that the existing stability analysis of shapes of minimal bending energy, as presented in, for example, Refs.\cite{milner1987dynamical,jaric1995vesicular}, does not directly apply to dynamic equilibrium since the latter is not known to minimize an energy functional. In particular, when the membrane relaxes from any non-axisymmetric perturbation of an equilibrium shape with $w\ne0$, it must dissipate kinetic energy. Therefore, under general shape perturbations, the system cannot relax to the same state and may find a close stationary state, transit to another branch, or relax to complete rest with $w=0$. A numerical illustration of the fluid deformable surface evolution from an oblate-biconcave shape towards prolate-dumbbell shapes with different symmetry axes can be found in Ref.\cite{krause2022numerical}.

\begin{rem}\rm  {The shape branches reported above were computed by trying different initial guesses in the algebraic solver. For example, the shapes in Fig.~\ref{fig8} resulted from setting the initial guess to be the perturbation of the unit sphere by $\tfrac1{10}S_2$, where $S_2$ is the 2nd spherical harmonic, and $\widehat V=0.95$. After the first shape was computed, we continued to ``move'' along the branch by gradually decreasing $\widehat V$ until the solver failed to converge starting from the previous shape as an initial guess.}        
\end{rem}

\subsection*{Larger $w$  and $c_\kappa=0$ cases.}

\begin{figure*}
	\begin{center}
		\begin{minipage}[c]{0.45\textwidth}
			\includegraphics[width=\textwidth]{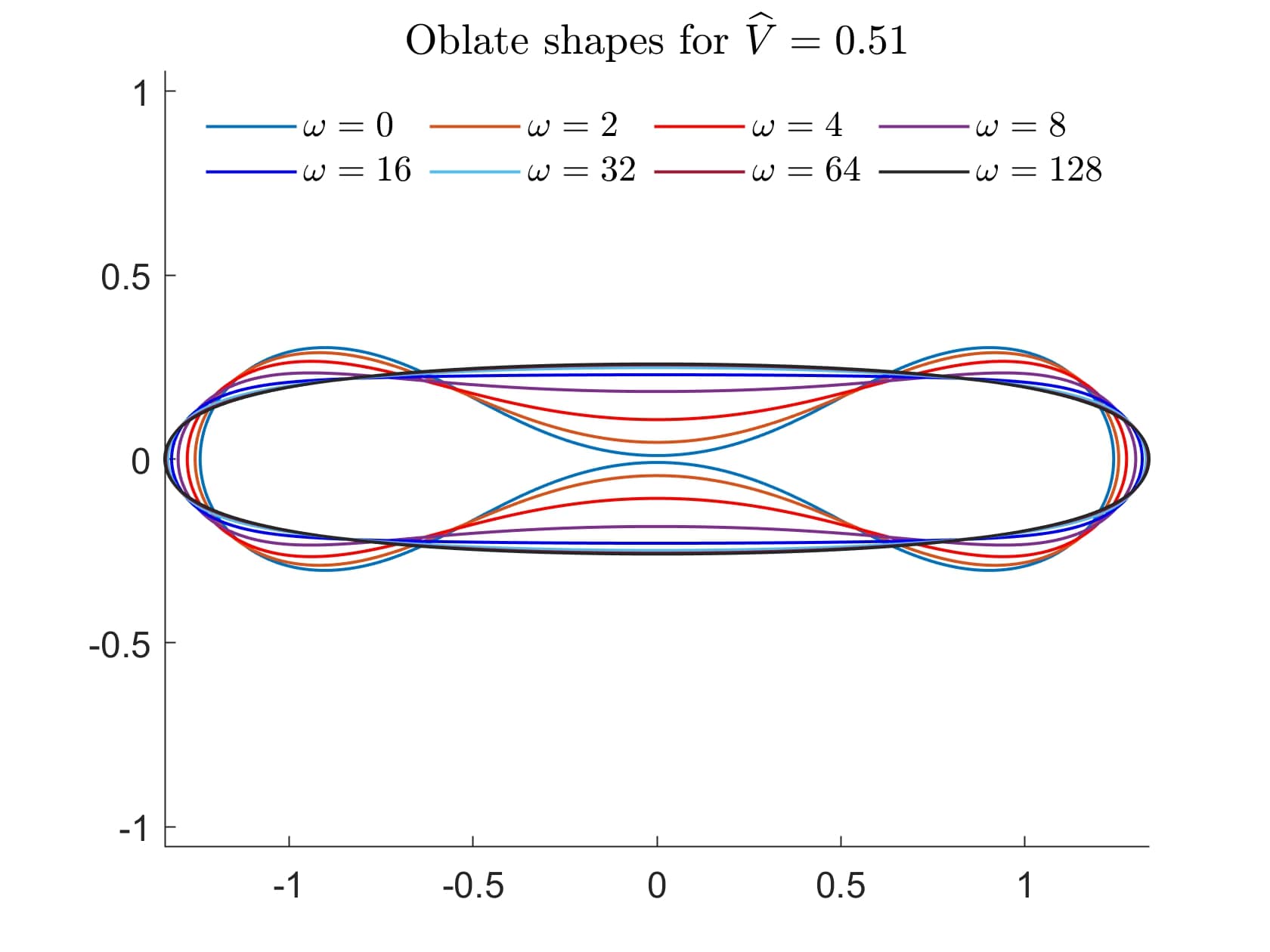}
		\end{minipage}
		\begin{minipage}[c]{0.5\textwidth} \small		
			\begin{tabular}{l|cccl}	
				$ w$ & 0 &2 & 4 &~~8  \\ \hline
				$ p_0$ &   4.10 &  0.69 & -10.3  & -57.6  \\
				$p^{\rm ext}$ &  (-0.016 &   -0.017 &   -0.018&   -0.016)$\times10^{3}$ \\
			\end{tabular}
			\vskip1ex
			\begin{tabular}{l|cccl}	
				$ w$ & 16 &32 &64 &~~128  \\ \hline
				$ p_0$ &   -251 &  -1036  & -4183 &   -16775\\
				$p^{\rm ext}$ &  (   0.011 &    0.148 &    0.728 &    ~~3.081)$\times10^{3}$ \\
			\end{tabular}
		\end{minipage}		
	\end{center}
\vskip-3ex
	\caption{\label{fig4}\small  Evolution of shapes with increasing $ w$ for $\widehat V=0.51$.}
\end{figure*}

To conclude this section, we will examine some shape transformations that occur when fluid inertia forces dominate over bending forces. Figure~\ref{fig4} depicts a branch of disc-like shapes for a reduced volume of $\widehat{V}=0.51$. The branch begins with a biconcave shape that solves equation \eqref{ShapeEq} for $w=0$, and continues with solutions for a sequence of increasing $w$. As $w$ becomes larger, we observe that the discs become less concave and eventually converge to a shape resembling an oblate ellipsoid.

\begin{figure*}
	\begin{center}
		\includegraphics[width=0.45\textwidth]{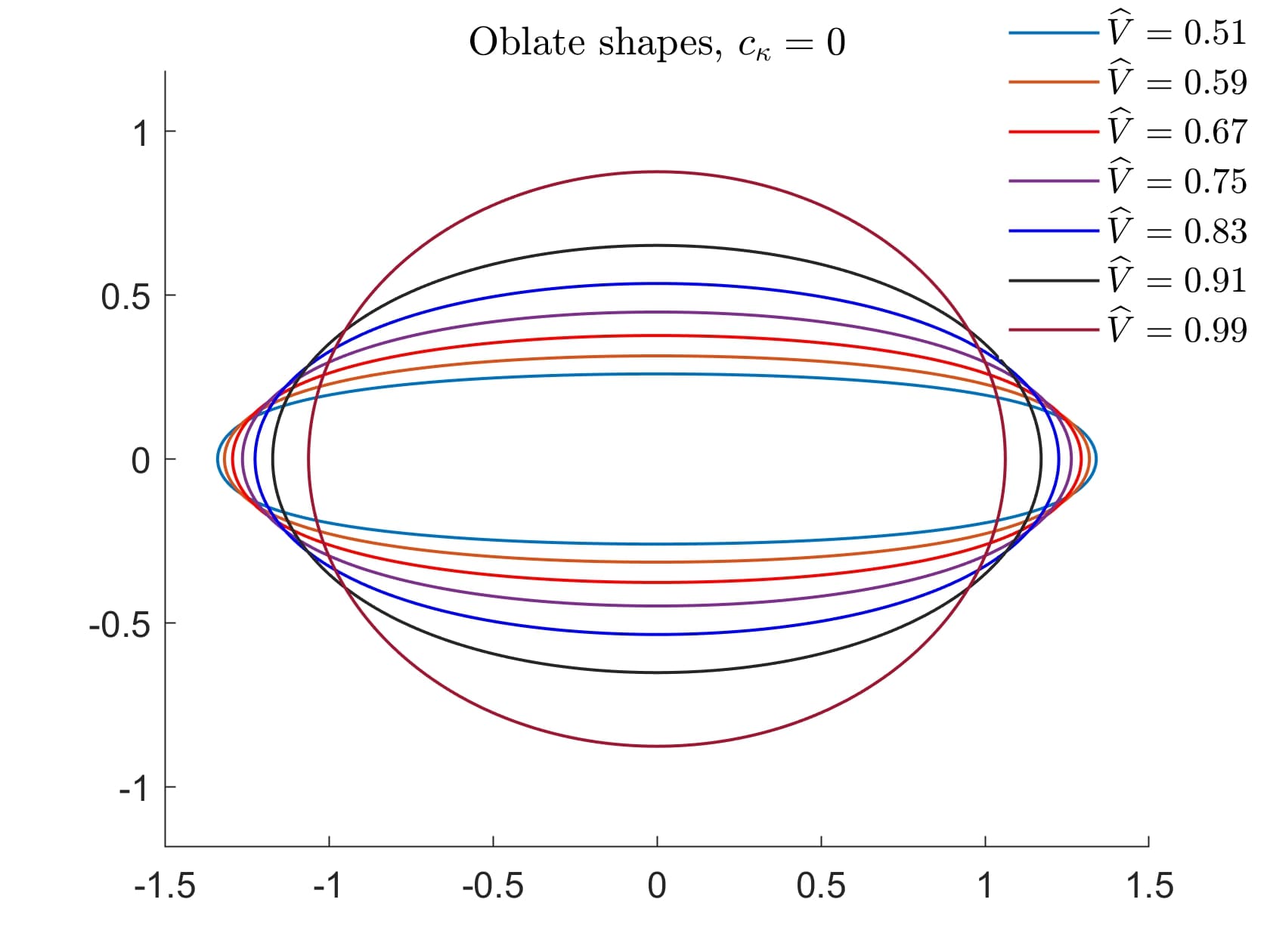}
		\includegraphics[width=0.5\textwidth]{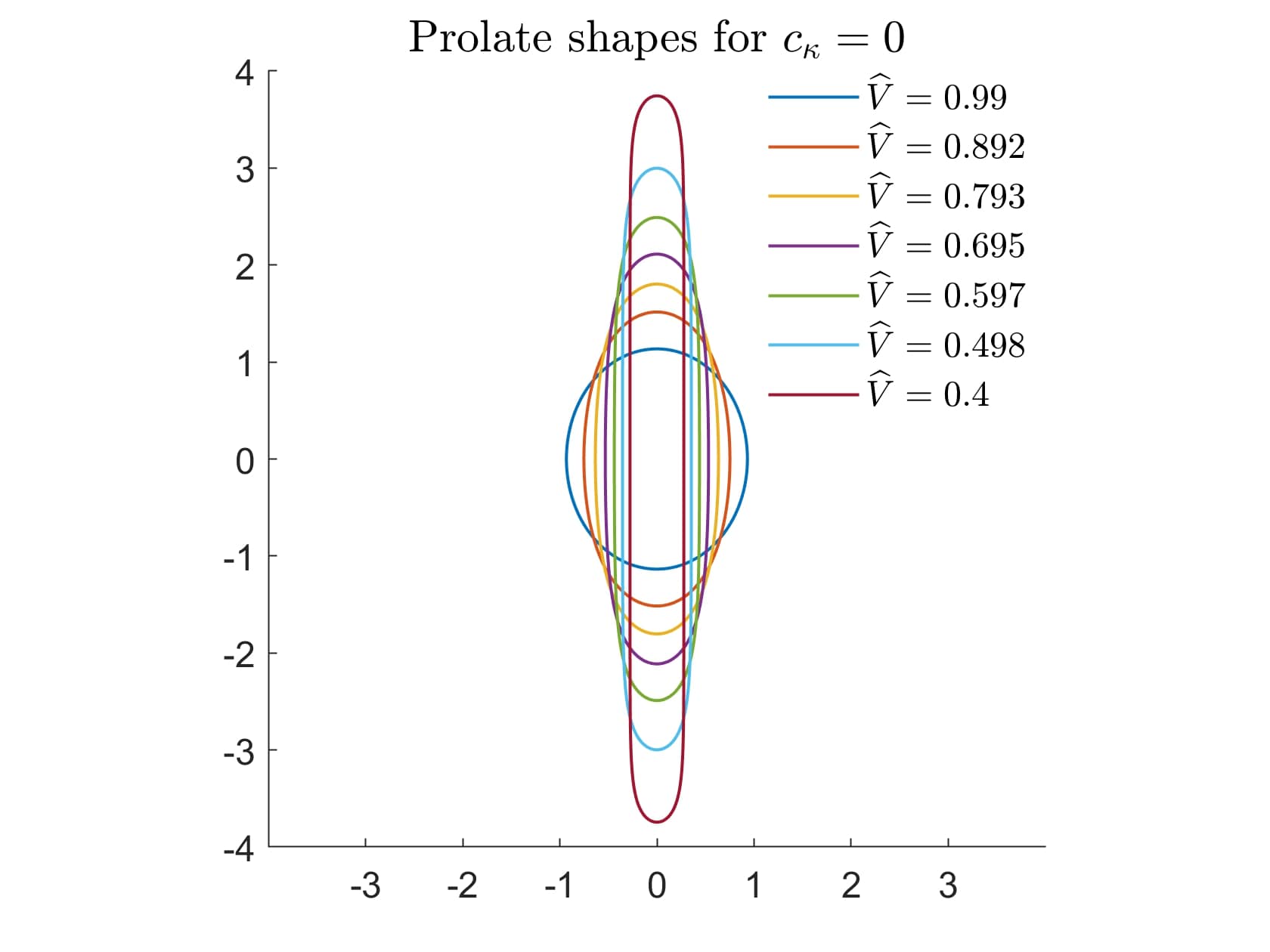}
	\end{center}
\vskip-3ex
	\caption{\label{fig10}\small Two branches of shapes for a pure fluid membrane, $c_\kappa=0$.}
\end{figure*}

If the  limit  (for $w\to\infty$) smooth surface exists, it solves the shape equation \eqref{ShapeEq} for the ``pure fluid'' case, in which the elastic forces are neglected by setting $c_\kappa=0$. {Mechanically this models a ``heavy'' fluid membrane such that inertia dominates over elasticity.}  We solved \eqref{ShapeEq} for $c_\kappa{=0}$ and found two branches of solutions consisting of oblate and prolate shapes, as shown in Fig.\ref{fig10}. In this limit case, we did not find any equilibrium states with concave or saddle shapes, as both principle curvatures were always positive. Additionally, we did not find any other solutions besides the two branches illustrated in Fig.\ref{fig10}. The coefficients $p_0$ and $p^{\rm ext}$ corresponding to the shapes in Fig.\ref{fig10} are reported in Table~\ref{tab1}.

\begin{table*}
	\small
	\begin{tabular}{l|ccccccc|ccccccc}	
		& \multicolumn{7}{c}{Oblate shapes} & \multicolumn{7}{c}{Prolate shapes}\\
		$\widehat{V}$& 0.51  &  0.59 &   0.67&   0.75&  0.83&  0.91&  0.99
		&0.40 &   0.498&   0.597  &  0.695 &0.793  &0.892  &0.99 \\ \hline
		&\multicolumn{14}{c}{$ w=0$, $c_\kappa=1$}\\
		$ p_0$  &  4.11&   4.64&    5.12&    5.55&    5.91&  6.17&  6.17&
		21.1  & 13.6  & 9.46  &  6.93  &  5.15 & 5.08& 5.74\\
		$p^{\rm ext}$ &  -16.1& -15.7&  -15.3&  -14.8&  -14.2&  -13.6&  -12.5&
		-105&  -54.5 & -31.7 & -19.9&  -13.0 & -11.4&  -11.6\\ \hline
		&\multicolumn{14}{c}{$ w=1$, $c_\kappa=0$}\\
		$ p_0$  &  -1.02&   -1.04&   -1.07&   -1.11&   -1.19&   -1.39&   -3.06&
		2.17  &  0.424  & 0.232  &  0.147  &  0.097   & 0.063  &0.038\\
		$p^{\rm ext}$ &  0.19&  0.26&    0.36&   0.50&  0.72&  1.21&  4.70 &
		-5.57  & -1.879  & -1.367 & -1.086 & -0.881 & -0.709   &-0.555 \\
	\end{tabular}
	\caption{\label{tab1}\small Surface tension coefficient $ p_0$ and osmotic pressure $p^{\rm ext}$ recovered for branches of oblate and prolate shapes for pure elastic and pure fluid shapes.}
\end{table*}

When seeking solutions to \eqref{ShapeEq} using the non-linear solver, we always used initial guesses that were symmetric with respect to the $xy$-plane. Therefore, asymmetric solutions are not reported in this study. However, we note that for the pure elastic model, asymmetric solutions are known to bifurcate from branches of symmetric shapes; see, for example, Refs.\cite{seifert1997configurations,jenkins1977static}, or the discussion in Sections~3.1.4,~3.4 of Ref.\cite{seifert1997configurations}. Therefore, asymmetric stationary solutions may also exist for $w\neq0$.

\section{Conclusions} \label{sec:concl}
The mechanical equilibrium of a fluid inextensible membrane with non-negligible mass is achieved through steady-state solutions of the surface Navier-Stokes equations coupled with an out-of-plane elasticity model. Assuming the Boussinesq-Scriven constitutive law for viscous stresses, we derived three conditions for membrane equilibrium, namely \eqref{cond1}, \eqref{Killing}, and \eqref{pressure}, which are independent of the elasticity model.

The second condition implies that there are only two possible scenarios: either the lateral motions of the membrane completely cease, or the equilibrium shape supporting a stationary flow is the surface with a Killing field. Accounting for a specific elasticity model leads to the shape equation. For elasticity models {with an} energy functional, the shape equation under the first scenario reduces to the optimality condition for the functional with area and volume constraints. For the second scenario, the shape equation represents a balance between the normal components of centrifugal, elastic, tension, and external forces. For axisymmetric surfaces, the equation can be efficiently parameterized and solved numerically.

Numerical studies using the simplest Helfrich elasticity model show that the equilibrium shapes depend on lateral motions {and may differ significantly from those known for $w=0$}. In particular, new branches of solutions appear. We also found some equilibrium states for a pure fluid membrane, which correspond to stationary solutions of the evolving-surface Navier-Stokes equations with no elastic forces and external forces given by a constant force acting in the normal direction (i.e., the constant osmotic pressure). Determining which of the computed equilibrium states are stable is an important open question that we leave for future research.

\begin{acknowledgments}
The author was supported in  part by the U.S. National Science Foundation under awards DMS-2011444 and DMS-1953535. It is a pleasure
to thank Robert Bryant and Gordon Heier for their help in understanding surfaces with Killing fields.
\end{acknowledgments}

\section*{Data Availability Statement}
A Matlab script used to generate the results presented in the paper can be obtained from the author upon a reasonable request.

\appendix
\section{}
Consider a smooth closed  $\Gamma$ embedded in $\mathbb{R}^3$. For a smooth vector field $\bv:\mathcal{O}(\Gamma)\to\mathbb{R}^3$
define
$
	\Gamma(t)=\{\by\in\R^3~|~\by=\bx(t,\bz),~\bz \in \Gamma \},
$
$t\in[0,\eps)$,
where the trajectories $\bx(t,\bz)$ solve the Cauchy problem
$
		\frac{d\bx}{dt} = \bv(\bx), \bx(0,\bz)=\bz\in\Gamma,
$
and a small $\eps>0$ such that $\Gamma(t)\in\mathcal{O}(\Gamma)$ for all  $t\in[0,\eps)$. Obviously, we have $\Gamma(0)=\Gamma$.
Applying the surface Reynolds transport theorem, one computes
\begin{equation}\label{aux696a}
\begin{split}
\left.\frac{dH}{d\Gamma}\right|_{\bv} &=\left.\left(\frac{d}{dt}\frac{c_\kappa}2\int_{\Gamma(t)}\kappa^2\, ds\right)\right|_{t=0}\\
&= \frac{c_\kappa}2\int_{\Gamma}\big(\dot{\kappa^2} +\kappa^2\divG\bv \big)\, ds\\
&= \frac{c_\kappa}2\int_{\Gamma}\big(2\kappa\dot{\kappa} +\kappa^2\divG\bv \big)\, ds.
\end{split}
\end{equation}
Let $d=d(t):\mathcal{O}(\Gamma)\to\mathbb{R}$ be a sign distance function for $\Gamma(t)$ which is smooth in the sufficiently small neighborhood  $\mathcal{O}(\Gamma)$. Then  $\bn=\nabla d$, $\kappa=\Div_\Gamma\bn$ are extensions of the normal vector field and mean curvature to  $\mathcal{O}(\Gamma)$.
We split $\bv$ into tangential and normal component:
\[
\bv=\bv_T+ v_N\bn.
\]
Since $\kappa$ is defined in a neighborhood of $\Gamma$, we can expand
\begin{equation}\label{aux719}
\dot{\kappa}=\tfrac{\partial\kappa}{\partial t}+\bv\cdot\nabla\kappa= \tfrac{\partial\kappa}{\partial t}+\bv_T\cdot\nabla_\Gamma\kappa+v_N(\bn\cdot\nabla)\kappa\quad\text{on}~\Gamma.
\end{equation}
Integration by parts along $\Gamma$ proves the identity
\begin{equation}\label{aux723}
2\int_{\Gamma}\kappa\bv_T\cdot\nabla_\Gamma\kappa\, ds=-\int_{\Gamma}\kappa^2\divG\bv_T\, ds.
\end{equation}
The identity  $\Div_\Gamma\bn=\kappa$ yields
\begin{equation}\label{aux723b}
\begin{split}
\divG\bv&=\div\bv_T+\divG(v_N\bn)\\
&=\div\bv_T+\bn\cdot\nabla_\Gamma v_N +v_N\kappa\\
&=\div\bv_T+v_N\kappa.
\end{split}
\end{equation}
Using \eqref{aux719}, \eqref{aux723} and \eqref{aux723b} in \eqref{aux696a} gives
\begin{equation}\label{aux696}
	\left.\frac{dH}{d\Gamma}\right|_{\bv}
	= \frac{c_\kappa}2\int_{\Gamma}\big(2\kappa(\tfrac{\partial\kappa}{\partial t}+v_N(\bn\cdot\nabla)\kappa) +\kappa^3 v_N \big)\, ds
\end{equation}
We assume that  the neighborhood $\mathcal{O}(\Gamma)$ is sufficiently small such that the closest point projection  $p:\mathcal{O}(\Gamma)\to\Gamma(t)$, $p(x,t)=x-d\bn$ is well defined. We then have
\begin{equation}\label{aux696b}
\frac{\partial d}{\partial t}=-v_N^e\quad \text{in}~\mathcal{O}(\Gamma)
\end{equation}
where $v_N^e(x,t)=v_N(p(x,t),t)$. With the help of \eqref{aux696b} and $\kappa=\Div_\Gamma \bn=\Div \bn=\Delta d$, we compute
\begin{equation}\label{aux703}
\begin{split}
\tfrac{\partial\kappa}{\partial t}&+v_N(\bn\cdot\nabla)\kappa\\
&=\Delta\tfrac{\partial d}{\partial t}+v_N\bn\cdot\nabla\Div\bn\\
&=-\Delta v_N^e+v_N\bn\cdot\nabla\Div\bn\\
&=-\Delta_\Gamma v_N+v_N\bn\cdot\nabla\Div\bn
\end{split}\quad \text{on}~\Gamma(t).
\end{equation}
Taking the divergence of the identity $\nabla\bn^2=0$, we get $0=\bn\cdot\Delta\bn+\nabla\bn:\nabla\bn$ implying that
$-\bn\cdot\Delta\bn=\mbox{tr}((\nabla\bn)^2)=\mbox{tr}(\bH^2)=\kappa_1^2+\kappa_2^2$. We use this  and $\Delta=\nabla\Div-\nabla\times\nabla\times$ to handle the last term in the right-hand side of \eqref{aux703}:
\begin{equation}\label{aux710}
\begin{split}
\bn\cdot\nabla\Div\bn&= \bn\cdot\Delta\bn+\bn\cdot(\nabla\times\nabla\times\bn)\\
&=-(\kappa_1^2+\kappa_2^2) +\bn\cdot(\nabla\times\nabla\times\bn)\\& =-(\kappa_1^2+\kappa_2^2).
\end{split}
\end{equation}
For the last equality we used $\nabla\times\bn=\nabla\times(\nabla d)=0$.
 Substituting \eqref{aux703}--\eqref{aux710} in \eqref{aux696}
we obtain
\begin{equation*}
\begin{split}
	\left.\frac{dH}{d\Gamma}\right|_{\bv}&
	= \frac{c_\kappa}2\int_{\Gamma}2\kappa(-\Delta_\Gamma v_N -(\kappa_1^2+\kappa_2^2)v_N) +\kappa^3 v_N \, ds\\
	&= \frac{c_\kappa}2\int_{\Gamma}2\kappa(-\Delta_\Gamma v_N -(\kappa^2 - 2K)v_N) +\kappa^3 v_N \, ds.
 \end{split}
\end{equation*}
Integration by parts yields the result in \eqref{bN}.


%
%

%




\section*{References}

\bibliography{literatur}

\end{document}